%% file: main.tex
\documentclass[twocolumn]{aastex63}
\pdfoutput=1 %for arXiv submission
\usepackage{amsmath,amstext,amssymb}
\usepackage[T1]{fontenc}
\usepackage{apjfonts} 
\usepackage{subfigure}
\usepackage{bm} % extends standard bold math
\usepackage{xcolor}
\usepackage{hyperref}
\usepackage{enumitem} % options for enumerate environment
\usepackage{soul}
\usepackage{comment}
\usepackage{graphicx}
\usepackage{tikz}
\usetikzlibrary{positioning}
\usepackage[normalem]{ulem}

\setstcolor{red}
\usepackage{lineno}
\graphicspath{{figures/}}

\newcommand{\sevn}{{\sc Sevn}}

\shorttitle{Probing the origin of the kilonova candidate GRB 230307A}
\shortauthors{C. R. Bom et al.}

\begin{document}

\title{Probing the origin of the kilonova candidate GRB 230307A:  analysis of host galaxy and offset}

\author[0000-0003-4383-2969]{Clecio R. Bom}
\affiliation{Centro Brasileiro de Pesquisas F\'isicas, Rua Dr. Xavier Sigaud 150, 22290-180 Rio de Janeiro, RJ, Brazil}
\author[0000-0003-1683-5443]{Davi C. Rodrigues}
\affiliation{Departamento de Física \& Núcleo Cosmo-Ufes, Universidade Federal do Espírito Santo. Av. Fernando Ferrari, 514, 29075-910. Vitória, ES, Brazil}
\affiliation{Centro Brasileiro de Pesquisas F\'isicas, Rua Dr. Xavier Sigaud 150, 22290-180 Rio de Janeiro, RJ, Brazil}
\author[0000-0002-0620-136X]{Arianna Cortesi}
\affiliation{Institute of Physics, Federal University of Rio de Janeiro, Av. Athos da Silveira Ramos 149, Rio de Janeiro, RJ 21941909, Brazil}
\affiliation{Observatório do Valongo/UFRJ, Ladeira do Pedro Antônio, 43 Centro, Rio de Janeiro, RJ 20080-090, Brazil}
\author[0000-0003-0293-503X]{Amanda E. Araujo-Carvalho}
\affiliation{Observatório do Valongo/UFRJ, Ladeira do Pedro Antônio, 43 Centro, Rio de Janeiro, RJ 20080-090, Brazil}
\author[0000-0002-2690-9728]{Daniel Ruschel-Dutra}
\affiliation{Universidade Federal de Santa Catarina, Campus Universitário
Reitor João David Ferreira Lima, 88040-900 Florianópolis, SC,
Brazil}

\author[0000-0003-0293-503X]{Giuliano Iorio}
\affiliation{Institut de Ciències del Cosmos (ICCUB), Universitat de Barcelona (UB), c. Martí i Franquès 1, 08028 Barcelona, Spain}
\author[0000-0003-3402-6164]{Luidhy Santana-Silva}
\affiliation{Centro Brasileiro de Pesquisas F\'isicas, Rua Dr. Xavier Sigaud 150, 22290-180 Rio de Janeiro, RJ, Brazil}

\author[0000-0002-5740-7747]{Charles~D.~Kilpatrick}
\affiliation{Center for Interdisciplinary Exploration and Research in Astrophysics (CIERA), Northwestern University, Evanston, IL 60208, USA}

\author[0000-0002-0056-1970]{Fabricio Ferrari}
\affiliation{Instituto de Matemática Estatística e Física, Universidade Federal do Rio Grande, 96203-900, Rio Grande, RS, Brazil}
\author[0000-0003-2127-2841]{Luis Lomel\'i-N\'u\~nez}
\affiliation{Observatório do Valongo/UFRJ, Ladeira do Pedro Antônio, 43 Centro, Rio de Janeiro, RJ 20080-090, Brazil}
\author[0000-0002-4130-636X]{Thomas Harvey}
\author[0000-0003-0519-9445]{Duncan Austin}
\author[0000-0003-1949-7638]{Christopher J. Conselice}
\author[0000-0003-4875-6272]{Nathan Adams}

\affiliation{Jodrell Bank Centre for Astrophysics, University of Manchester, Oxford Road, Manchester M13 9PL, UK;}
\author[0000-0001-9672-0296]{Roberto Cid Fernandes}
\affiliation{Universidade Federal de Santa Catarina, Campus Universitário
Reitor João David Ferreira Lima, 88040-900 Florianópolis, SC,
Brazil}

\correspondingauthor{Clecio R. Bom}
\email{debom@cbpf.br}

\begin{abstract}
We investigate the host galaxy of the long gamma-ray burst GRB 230307A, which is associated with a kilonova candidate likely produced by a binary neutron-star (BNS) merger. The transient occurred at a projected offset of $\sim 40 \mathrm{kpc}$ from its host. We consider two explanations for this large distance: (i) NSs that merge inside a remote globular cluster, or (ii) a BNS that formed in the disk whose orbit was strongly modified by the NS natal kicks.
Using JWST data and comparisons with known globular clusters, we show that a globular-cluster origin is unlikely.
Using JWST and MUSE data, we derive the host galaxy morphology, stellar mass, estimate the atomic gas (HI+He) contribution, and the host rotation curve. Assuming an NFW halo and applying Bayesian inference, we obtain a mass model for the host. From this model, we compute the time required for a disk-formed BNS, with a given natal kick, to reach the observed offset while marginalizing over uncertainties and over the initial position in the disk.
We compare these results with BNS-merger simulations from a population-synthesis code combined with stellar evolutionary tracks, which provide the coalescence time and kick velocity for each realization. 
The two approaches have an overlap in the kick-time diagram, but only 0.1\% of the simulated systems is compatible with the galaxy-mass model.
This indicates that a disk origin is possible, but requires fine-tuned conditions for the kilonova to occur at such a large distance from the host galaxy.
\end{abstract}


\section{Introduction}

The GRB230307A, an extremely bright and long  Gamma Ray Burst (GRB), was detected by the Gamma-ray Burst Monitor (GBM) \citep{2023GCN.33407....1D}. The extensive analysis of the afterglow spectra revealed that they are consistent with Lanthanide production, likely associated to a kilonova \citep{2024Natur.626..737L, 2024Natur.626..742Y}. Kilonova events, (i.e., explosions due to neutron stars or neutron star-black hole mergers) are rare astrophysical phenomena associated to the production of gravitational waves and with GRBs.

Here we focus our attention on the most likely host galaxy, which is at $z=0.0647$, corresponding to a distance of 291 Mpc using $H_0$ from \citet{Freedman_2019}, and has a considerably large projected offset of $\sim 40$~kpc from the GRB \citep{2024Natur.626..737L, 2024Natur.626..742Y}.
For simplicity, this host-galaxy candidate, WISEA J040318.87-752255.7, is here either referred as ``host'' or G1 galaxy, following \citet{2024Natur.626..737L, 2024Natur.626..742Y}.
The main features of G1 are similar to those of the host of GRB 211211A \citep{troja2022}, also associated with a possible kilonova from a long GRB.
Similarities include the low mass and star formation rate, as well as relatively  old stellar population, suggesting that some long-duration GRBs originate in such environments.

In this contribution, we analyze G1 to assess the probability of finding the kilonova at its observed position. We use JWST data to determine the host galaxy’s morphology and MUSE observations to derive its rotation curve, mass assembly history, ages, and metallicity. The stellar mass is inferred from SED fitting combining MUSE spectra with JWST photometry, following \cite{Cappellari2023}. The atomic gas (HI + He) contribution is estimated. A mass model is constructed using the aforementioned baryonic data, a Navarro-Frenk-White (NFW) dark matter halo \citep{Navarro:1996gj}, and Bayesian inference. To model the SNe natal kick distribution, we use the archival dataset of Stellar EVolution N-body \citep[\sevn;][]{iorio_2023} for simulated BNS population.

%%%%%%%%%%%%%
\section{Host Photometry and  Morphology} \label{sec:morphology}

\subsection{JWST photometry and photometric redshift determination}

We made use of JWST observations of the host galaxy of GRB230307A from The Mikulski Archive for Space Telescopes (MAST), which was obtained through JWST GO programs  4434 (PI Levan) and 4445, (PI Levan). These two epochs provided NIRCam imaging in the F070W, F115W, F150W, F277W, F356W and F444W filters. We reduced the raw data products using a modified version of the official JWST pipeline to create a multi-epoch mosaic in the filters observed in both epochs. The changes to the standard pipeline include a $1/f$ correction, improved artifact removal, and a modified background subtraction. We astrometrically align the imaging to GAIA DR3 \citep{GAIADR3} and pixel-align the images to F277W. This pipeline is described in detail in \cite{adams2024epochs, conselice2025epochs, Harvey2024}. 

Our catalogues were produced using the {\tt galfind}\footnote{\url{https://github.com/duncanaustin98/galfind}} python package; the methodology is outlined below. We use {\tt SExtractor} to perform source detection and measure aperture and isophotal photometry for all sources in the image \citep{sextractor}. As a detection image we use an inverse-variance weighted stack of the F277W, F356W and F444W images. We measure aperture photometry in $0.32, 0.5, 1, 1.5$ and $2$ arcsecond diameter apertures. We perform aperture corrections to all photometry measurements using simulated PSF models obtained through {\tt webbpsf} \citep{perrin2014updated}. 

We calculate localized photometric uncertainties for each aperture for each galaxy from the normalized median absolute deviation (NMAD) of the nearest 200 empty circular apertures of the same size which are randomly positioned in `empty' regions of the image (defined as being 1 arcsec away from occupied pixels in the segmentation map).

We estimate photometric redshifts for all galaxies in the field using {\tt eazy-py} \citep{Brammer_2008} and the standard set of basis {\tt fsps} templates \citep{conroy2010fsps}. We set a minimum flux error of 5\% to account for residual errors in flux calibration, zero-points and the templates themselves. 

We have combined the JWST images available from the two runs in order to create  final images with higher S/N, in the six available bands. These images have been used to obtain photo-z of all the objects in the field, in particular looking for neighboring galaxies, see Appendix \ref{app:photometry}.

\subsection{Kilonova within a Globular Cluster} \label{sec:globular}

One possible formation channel for BNS
systems could be hierarchical formation through globular clusters (GCs)~\citep[GCs;][]{Ivanova08},  given that GCs can be found in the halos of their galaxies at separations up to $\approx$100~kpc \citep[e.g., in][]{Lim24}, supporting a large offset. Therefore, we explore the possibility that the progenitor system of GRB\,230307A exploded within such a system. The depth of the {\it JWST}/NIRCam observations are such that luminous GCs, with typical absolute magnitudes near 1~$\mu$m from -5 to -11~mag \citep{Jordan07}, would be detected at the site of the GRB, and thus if the system formed and remained bound to the gravitational potential of its parent GC we would observe a counterpart in deep imaging.

\begin{figure}[ht]
    \includegraphics[width=0.49\textwidth]{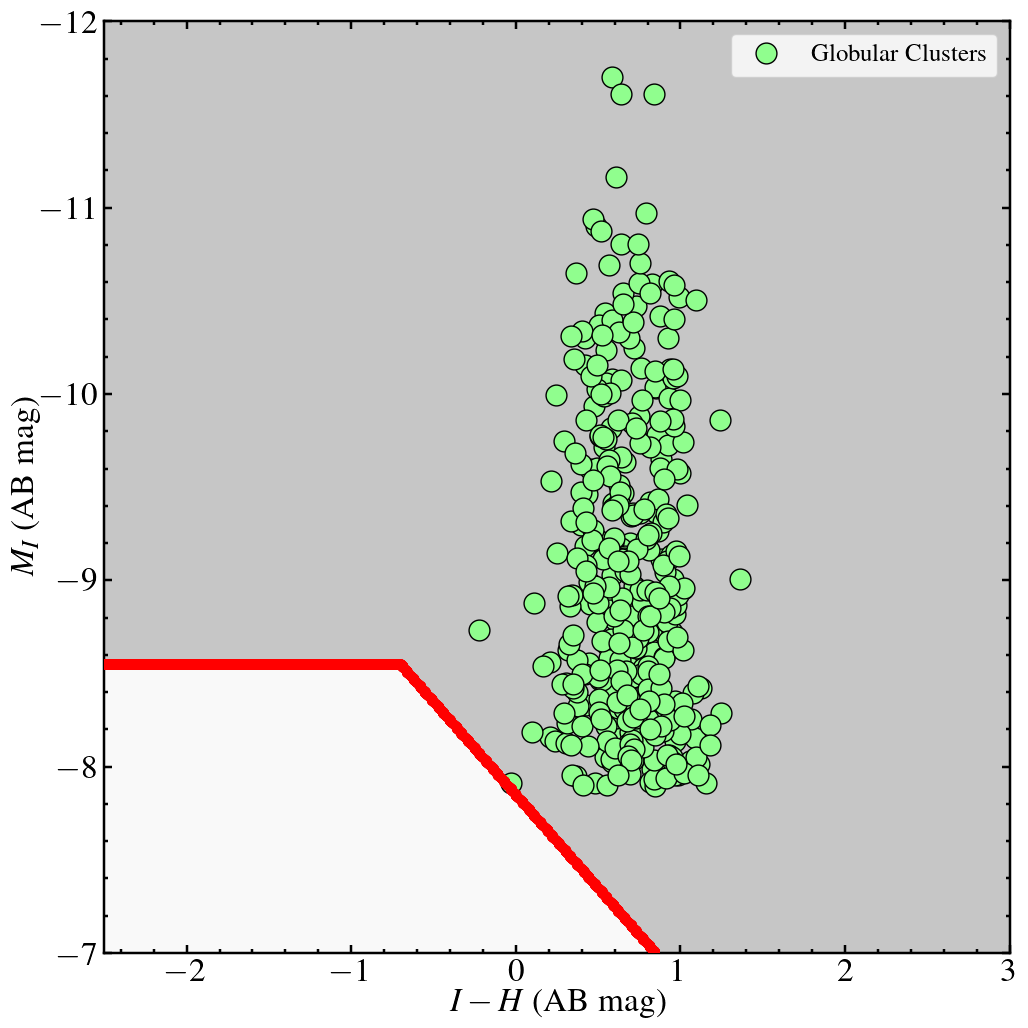}
    \caption{A color-magnitude diagram of a sample of globular clusters from \citet{Blakeslee12} in $I$- and $H$-band compared with limits on a counterpart to GRB\,230307A in F070W and F150W.  The red line corresponds to our joint detection threshold to point sources in F070W and F150W, with the grey region containing detected and the white region undetected point sources.  We rule out nearly all of the globular cluster luminosity function, implying that it is extremely unlikely that the progenitor of GRB\,230307A formed in situ in a globular cluster.}\label{fig:gclf}
\end{figure}

We note that the counterpart to GRB\,230307A was constrained in NIRCam imaging to depths of F070W$>$29.0~mag and F150W$>$28.6~mag\footnote{We report all photometry on the AB magnitude system.} by \citet{2024Natur.626..737L}.  
Interpreting these limits as comparable to $I$ and $H$-bands, respectively, we follow  the analysis in \citet{charle_gw170817} to constrain the fraction of the GC luminosity function \citep[GCLF; e.g., from][]{Lee18} to determine the likelihood that GRB\,230307A could have formed within a GC (Figure~\ref{fig:gclf}).  At the assumed distance of 291~Mpc, we can rule out the entire GCLF derived from galaxies such as NGC\,1399 \citep{Blakeslee12} from an empirical distribution approach (since there are no observed GC in this region).  However, the region of the $I-H$ versus $M_I$ plane not probed by JWST lies adjacent to the region populated by globular clusters, with at least one GC nearly overlapping this region. Consequently, a strictly zero-probability assignment based on the empirical distribution alone would place stronger constraints than supported by the data.
    
To estimate the order of magnitude of the probability that a globular cluster could reside within the unobserved JWST region, we model the data using a Gaussian KDE to infer a smooth underlying distribution. For this purpose, all standard bandwidth choices considered (Silverman, Sheather-Jones, and Scott) yield consistent results, implying a probability of $(0.1$--$0.3)\%$ for the presence of a globular cluster in the unobserved region. As expected, considering that the sample has 400 GCs with one nearly inside the excluded region, the probability is quite low, but not zero. Larger GCs samples can improve this probability estimation.

\subsection{Host galaxy morphological parameter determination}

To retrieve the host galaxy morphological parameters, we have first used the code {\sc Morfometryka} \cite[MFMTK,][]{morfometryka} to perform photometry and create masks of objects that might affect the fit.
Three small galaxies fall within the host galaxy isophotes, but their photo-z is consistent with $z\simeq1$ (see Appendix \ref{app:photometry}). It is necessary to mask them  to obtain the correct value of the galaxy integrated photometry, as well as to recover its photometric profile. 
MFMTK only needs the galaxy image and its PSF to run. We selected the F356W band, which better traces the old stellar population, and we created the PSF using PSFex \citep{bertin_psfex}. The output of MFMTK, \citep{Conselice_20}, are shown in Appendix \ref{app:photometry}.
 
We used the recovered mask, center determination, axis ratio and effective radius from MFMTK as initial parameters for GALFIT \citep{galfit_peng2002} to perform a single exponential profile fit \citep[e.g.][]{vanderKruit:2011vt, 2023A&A...677A.117S}. Figure \ref{fig:kine} (middle panel) shows that these methods yield surface brightness profiles which are compatible with a single exponential profile.\footnote{The data show no clear evidence for down-bending (Type II) or up-bending (Type III) profiles \citep{2023A&A...677A.117S}.} 
Since GALFIT directly fits the 2D observational data, henceforth we use its $h$ value. The smaller $h$ value from MFMTK will slightly reduce the likelihood of finding a kilonova far from the host galaxy. Further details are in Appendix \ref{app:photometry}. The exponential profile provides a convenient, usual and transparent approximation that will be used in the next sections.

\section{Host Kinematics}\label{sec:kinematics}
Although the available MUSE data is not deep enough for accurate estimates of the stellar kinematics with high spatial resolution, G1 has intense emission lines from the ionized gas in the ISM.
These lines were fit by Gaussian profiles sharing the same kinematical parameters, namely line-of-sight velocity and velocity dispersion.
The kinematic properties discussed below were derived from fits of prominent emission lines near the H$\alpha$ line: 
[\ion{N}{2}]$~\lambda6548${\AA},
H$\alpha$,
[\ion{N}{2}]$~\lambda6583${\AA},
[\ion{S}{2}]$~\lambda6716${\AA},
and [\ion{S}{2}]$~\lambda6731${\AA}.
The actual fits were carried out by IFSCube \citep{2020zndo...3945237R,2021MNRAS.507...74R}, which is a {\sc python} package for the processing and analysis of astrophysical datacubes, with an emphasis on optical spectra.

Since the velocity maps of the gas do not show clear signs of dynamical disturbances they are a reasonable proxy for the kinematics of the stars as well.
In order to find the kinematical center of the galaxy we fit a 2D model of a rotating disk \citep{bertola_testing_1991} to the velocity map of G1.
This revealed a slight misalignment between the optical and kinematic cores of about $0.2^{\prime\prime}$, which is the length of one spatial pixel and less then the FWHM of the PSF for the MUSE data.
Another result from this fit is the kinematical position angle, which was found to be $43^\circ$. 

\begin{figure*}[!]
\begin{center}
\includegraphics[width=\linewidth]{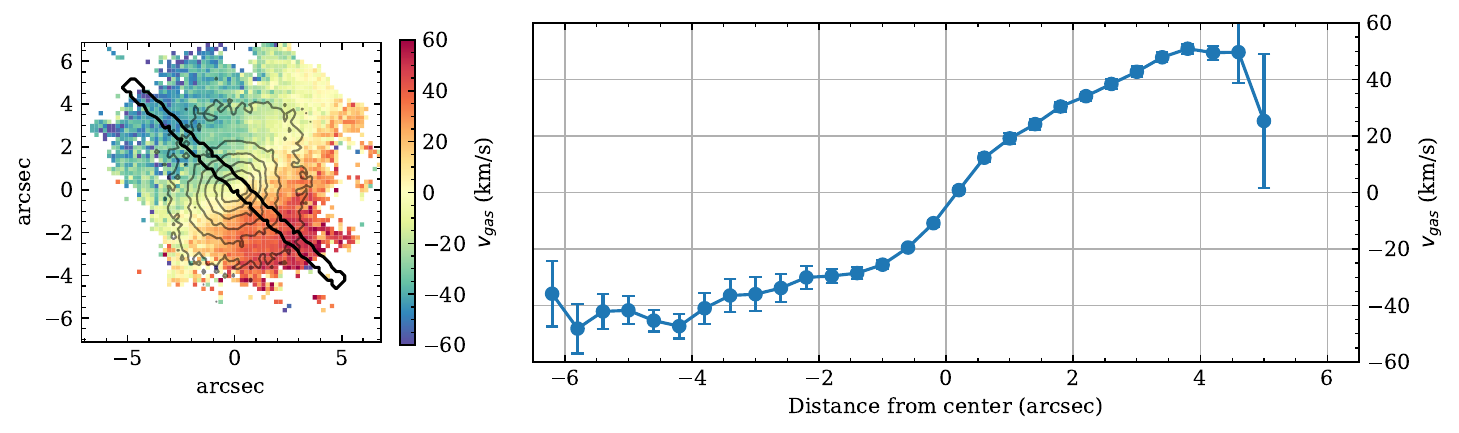}
\end{center}
\caption{
    \emph{Left:} Velocity map of the H$\alpha$ emission (see Sec.~\ref{sec:kinematics}).
    The solid black line represents the path of virtual extractions.
    A clear rotation pattern can be discerned in this map.
    \emph{Right:} The rotation curve extracted from the map on the left with error bars indicating the $1\sigma$ interval of uncertainty.
}
\label{fig:kine}
\end{figure*}

To find the rotation curve, which is the subject of the next section, the 2D velocity map is reduced to a 1D velocity profile.
This profile was extracted from the velocity maps by averaging the velocity within virtual circular apertures along a straight line that crosses the kinematical center and has the same position angle of the rotating disk model.
Each aperture has a radius of $0.4^{\prime\prime}$ matching the spatial FWHM of the observations.
The results from this procedure are shown in Fig. \ref{fig:kine},
with the line-of-sight velocity map in the left panel, and the velocity curve along the kinematical axis on the right panel.
Error bars in the velocity curve represent the standard deviation estimated from 30 bootstrap iterations, each obtained by randomizing the spectral flux density according to its variance under the assumption of Gaussian errors.

\begin{figure*}[!]
\begin{center}
\includegraphics[width=0.95\linewidth]{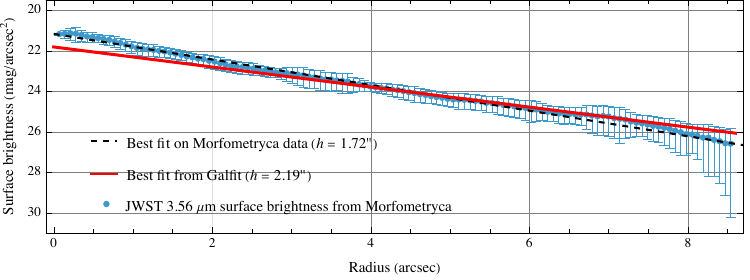}
\includegraphics[width=\linewidth]{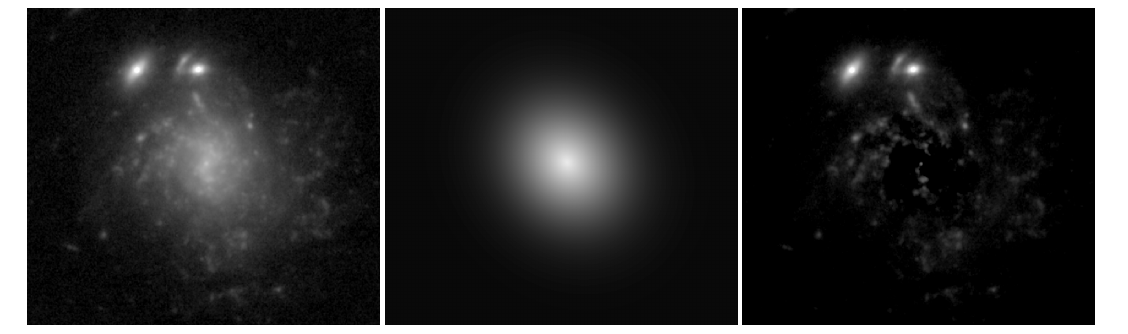}
\end{center}
\caption{
     \emph{Upper:} Surface brightness profile obtained extracting the galaxy light in elliptical bins, following the galaxy axis ratio, of increasing radius, shown as blue dots with error bars. The black dashed line represents the 1D fit of these data, assuming an exponential profile. 
     The red line shows the output of the 2D GALFIT for an exponential profile, which leads to a larger disk scale length ($h$) than the previous case.  \emph{Bottom:} GALFIT fit: left image, middle model, and right residuals.
}
\label{fig:SBB}
\end{figure*}

\section{Rotation curve and galaxy mass models} \label{sec:rotCurve}

From the previous section results, we infer the line-of-sight velocity for the approaching and receding sides. Following \citet{2009A&A...493..871S}, these two sides are combined and corrected for inclination, as detailed in the Appendix \ref{app:rotDerivation}. From the latter, the rotation curve, denoted as $V_{\rm rot}(R)$, is found. It does have a dependence on the inclination, which will be considered as a free variable with a prior.  The mass model detailed in the next subsections decomposes the galaxy into a stellar disk, a gaseous disk (mainly constituted by atomic hydrogen and helium) and a dark matter halo. 

\subsection{Stellar component model} \label{sec:stellarModel}

To model the stellar contribution to the rotation curve, we consider a thin exponential disk \citep{vanderKruit:2011vt}. As shown in Fig.~\ref{fig:kine}, this simple and commonly used approximation is in good agreement with this galaxy, up to the observational limit. This stellar model can be parameterized by the stellar mass $M_\star$ and the disk-scale length $h$. For $M_\star$, we derive here $\log_{10}(M_\star/M_\odot) = 9.68 \pm 0.10$ from SED fitting (Appendix~\ref{sec:StellarMass}). For $h$, we use the 3.56 $\mu$m  band data \citep{Meidt:2014mqa} and GALFIT to fix it as $h = 2.19'' = 2.90 \mbox{ kpc}$ (Sec.~\ref{sec:morphology}).  

\subsection{Gas component model} \label{sec:gasModel}

Following \citet{2002A&A...390..863S} and \citet{2016AJ....152..157L}, we approximate the HI profile as a thin exponential disk with a scale length ($h_{\rm HI}$) twice that of the stellar disk, i.e., $h_{\rm HI} = 2 h$. Such approximations are here used since, to the best of our knowledge, direct observations of the atomic hydrogen (HI) distribution for the host galaxy are not currently available. To determine the total HI mass, we use the empirical correlation 
\begin{equation} \label{eq:MhiRhi}
    \log M_{\rm HI} = 1.87 \log R_{\rm HI} + 7.20 \, ,
\end{equation}
where $R_{\rm HI}$ is the radius at which the HI density reaches 1 $M_\odot \, \mbox{pc}^{-2}$. 
Therefore, for the HI surface density $\Sigma_{\rm HI}(R) = \Sigma_{0, \rm HI} e^{-R/h_{\rm HI}}$, one writes
\begin{equation}
    R_{\rm HI} = 2 h  \ln \left(\Sigma_{0,\rm HI}\,  10^{-6}\right) \, ,
\end{equation}
with $\Sigma_{0, \rm HI}$ in $M_\odot$ kpc$^{-2}$ units. Hence, Eq.~\eqref{eq:MhiRhi} can be numerically solved for $\Sigma_{0, \rm HI}$, leading to
\begin{align}
    M_{\rm gas} = 1.33 M_{\rm HI} = 1.83 \times 10^9 M_\odot\, , 
\end{align}
where $M_{\rm gas}$ stands for HI and atomic He total mass. With these estimates, the gas contribution to the circular velocity is found. 

\bigskip
\subsection{Dark matter halo}

The Navarro-Frenk-White (NFW) dark matter halo is a simple and effective parametrization that is derived from N-body simulations \citep{Navarro:1996gj, MoBoschWhite2010} whose density profile is given by
\begin{equation}
    \rho_{\rm NFW} = \frac{\rho_s}{\frac{r}{r_s} \left( 1 + \frac{r}{r_s}\right)^2} \, .
\end{equation}
It is also commonly parametrized by the concentration $c$ and $r_{200}$, which satisfy $c = r_{200}/r_s$ and $M_{\rm NFW}(r_{200}) r_{200}^{-3} = 200 \rho_c$, where $M_{\rm NFW}(r)$ is the interior mass of the NFW halo and $\rho_c = 3 H_0^2/(8 \pi G)$. It is well known that the two NFW parameters are correlated, and such correlation will enter in the complete analysis as a prior.  

\bigskip
\subsection{Total mass model}

The total model rotational velocity in the $z=0$ plane is found from the Newtonian potential addition of each component, and it is given by 
\begin{equation} \label{eq:Vmodel}
  V^2_{\rm M}(\delta_\star, r_{200}, c) = \delta_\star V^2_\star + V^2_{\rm gas} + V^2_{\rm NFW}(r_{200}, c)\, .
\end{equation}
In the above, $\delta_\star$ parametrizes the uncertainty on the stellar mass, with $\delta_\star = 1$ being the most probable case. The analytical expressions for $V^2_\star$ and $V_{\rm gas}$ (since both are exponential disks) are well known \citep{BinneyTremaine} and are used in our codes. 

\bigskip
\subsection{The likelihood and the priors}

Denoting the collective parameters as $\Theta$, with $\Theta = (\delta_\star, \log c, \log r_{200}, \Delta i)$, where $\Delta i$ is the inclination deviation with respect to the most probable value, and denoting the rotation curve data as $D$, the posterior distribution is $f(\Theta|D) \propto f(D|\Theta) f(\Theta)$, where $f(\Theta)$ are the priors and $f(D|\Theta) = {\cal L}(\Theta)$ the likelihood \citep[see][]{Marra:2020sts}.  

Since the observational data of the rotation curve have Gaussian distribution, we use ${\cal L} \propto e^{- \chi^2/2}$, with
\begin{equation}
  \chi^2 = \sum_{j=1}^N \left(\frac{V_M(\delta_\star, r_{200}, c, R_j) - V_{\rm rot}(R_j, \Delta i)}{\sigma_j(\Delta i)}\right)^2 \, ,
\end{equation}
where $N$ is the total number of rotation curve data points, $R_j$ is the radius of each observed $V_{\rm rot}$ data point, and $\sigma_j$ the corresponding uncertainty. $V_{\rm rot}$ and  $\sigma_j$ depend on the inclination variation ($\Delta i$): $ V_{\rm rot}(R, \Delta i) =  V_{\rm los}(R)/\sin(i_0 + \Delta i)$ \citep[see e.g.,][]{Marra:2020sts},
with $V_{\rm los}$ the line-of-sight velocity and $i_0 = 0.59$ rad (i.e., $i_0 = 33.9^\circ$) is the central value of the inclination.

The priors are given by 
\begin{align}
    & f(\log \delta_\star) = {\mathcal N}(0, \, 0.10^2) \, ,\nonumber \\
    & f(\Delta i) = {\mathcal N}(0, \, 0.04^2) \, ,\nonumber \\
    & f(\log c ) =  {\mathcal N}( C_{\rm M08}, \, 0.11^2) \, , \label{eq:NFWpriors}\\
    & f\left[\log \left({r_{200}}/{{\rm kpc}} \right) \right] = {\mathcal U}(0,\,  4)  \nonumber \, ,
\end{align}
with $f(\Theta) = f(\log \delta_\star) f(\Delta i) f(\log c ) f\left[\log \left({r_{200}}/{{\rm kpc}} \right) \right]$.
In the above, $\mathcal U(a,\, b)$ is the uniform distribution in the interval $(a, b)$ (i.e., a flat prior), and  $\mathcal N(\mu, \, \sigma^2)$ is the normal distribution with central value $\mu$ and dispersion $\sigma$. In particular, $0.04$ in the above is the 1$\sigma$ uncertainty for the inclination in radians and we adopt  0.10 dex as the 1$\sigma$ uncertainty for the stellar mass correction $\delta_\star$ (which is a more conservative uncertainty than the formal and much smaller one found from SED fitting). The constant $C_{\rm M08}$ is 
\begin{equation}
    C_{\rm M08} = 0.830 - 0.098 \log\left(M_{200}/[10^{12} h^{-1} M_\odot]\right) \, .
\end{equation}
$C_{\rm M08} = \log c$ corresponds to the most probable $\log c$ value, as derived by  \citet{Maccio:2008pcd}, with a dispersion of 0.11 dex. 

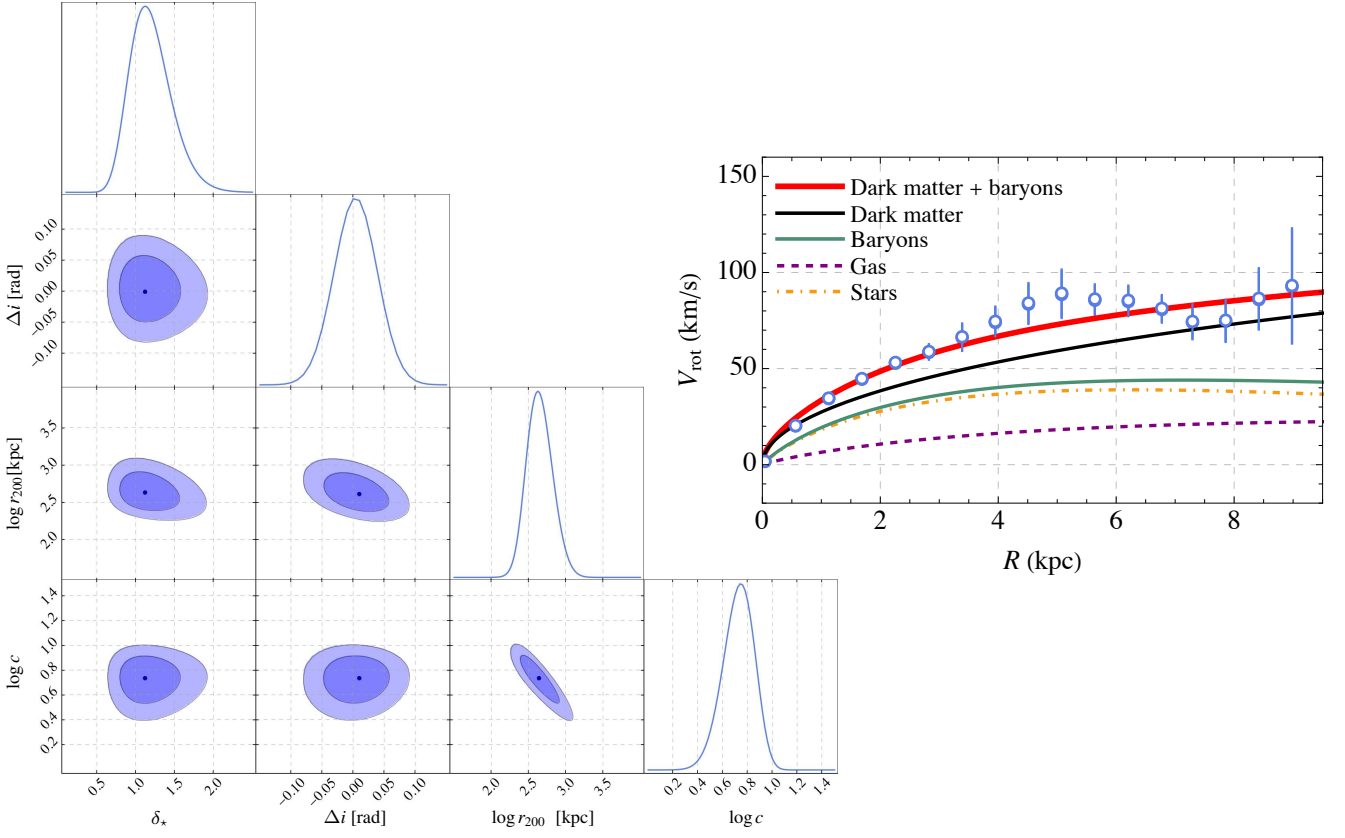
\begin{figure*}[ht]
        \begin{tikzpicture}
            % Horizontal
            \node (fig) at (0,0)  {\includegraphics[width=.60\linewidth]{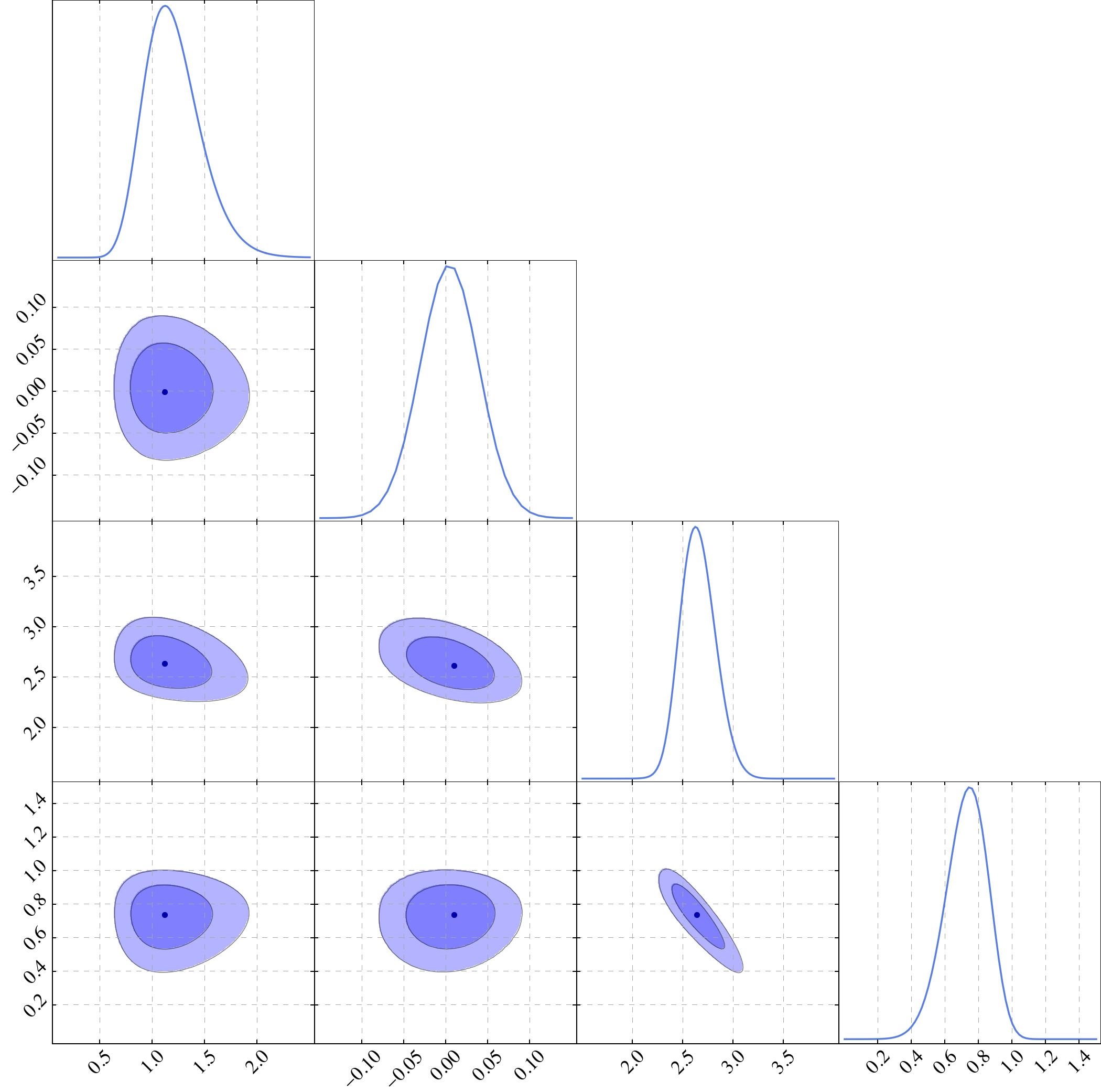}}; 
            \node[below=of fig,node distance=0cm, yshift=1.2cm, xshift=-3.6cm] {\scriptsize $\delta_\star$};
            \node[below=of fig,node distance=0cm, yshift=1.2cm, xshift=-1.0cm] {\scriptsize $\Delta i \mbox{ [rad]}$};
            \node[below=of fig,node distance=0cm, yshift=1.2cm, xshift=1.5cm] {\scriptsize $\log r_{200} \;$ \mbox{\rm [kpc]}};
            \node[below=of fig,node distance=0cm, yshift=1.2cm, xshift=4.2cm] {\scriptsize $\log c$};
            % Vertical
            \node[left=of fig,node distance=0cm, rotate=90, yshift=-1cm, xshift=-3.2cm] {\scriptsize $\log c$};
            \node[left=of fig,node distance=0cm, rotate=90, yshift=-1cm, xshift=-0.3cm] {\scriptsize $\log r_{200} \mbox{[kpc]}$};
            \node[left=of fig,node distance=0cm, rotate=90, yshift=-1cm, xshift=2cm] {\scriptsize $\Delta i \mbox{ [rad]}$};
            % Right figure
            \node (fig) at (7.6,0.5)  {\includegraphics[width=.48\linewidth]{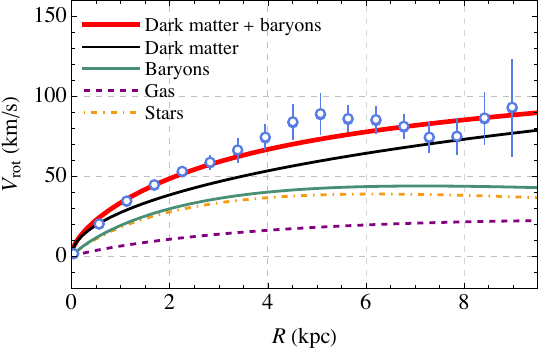}};
        \end{tikzpicture}
\caption{\textit{Left.} Marginalized 2D and 1D sections of the posterior distribution for the fitted parameters. In the 2D regions, the contours indicate the 1$\sigma$ and 2$\sigma$ credible regions, and the black dot marks the mode of each marginalized case. \textit{Right.} The best-fit (MAP) rotation curve with a NFW dark matter halo ($\delta_\star = 1.10, \; \Delta i = 0.01, \;  \log c = 0.75, \; \log (r_{200}/\mbox{kpc}) = 2.64$). }
\label{fig:nfw_rot_curve}
\end{figure*}

For the best-fit model parameters, we use the  maximum a posteriori (MAP) estimate. In Fig.~\ref{fig:nfw_rot_curve}, we show the resulting rotation curve and state the corresponding MAP parameter values. The credible parameter regions are inferred from highest density region (HDR) of the posterior distributions. Further details are in  Appendix \ref{app:rotDerivation}.

%%%%%%%%%%%%%%%%%%
\section{Natal kicks and the offset} \label{sec:natalKicks}

Here we present our main results. It is assumed that the GRB230307A is a kilonova generated by the collision of two NSs, since this scenario maximizes the velocity of the BNS system. All the kick values refer to the center of mass kick. As explained below, the analysis is based on a single effective kick. 
 
\subsection{The radial position probability density} \label{sec:offsetStellar}

In a spiral galaxy, a binary system of massive stars capable of producing two NSs is expected to form in the galactic disk, where most of the stellar mass resides and star formation is most active.

For an exponential stellar disk with $h$ as the disk-scale length, the probability density to find a star in a given disk radius $R$ is 
\begin{equation} \label{eq:pdfR}
    f(R) = \frac{1}{h^2} e^{-R/h} R \, ,
\end{equation}
whose maximum is at $R=h$. 

The probability $P(d)$ of finding a star at a radius $R \geq d$ in the disk is given by integrating \eqref{eq:pdfR} from $R=d$ to infinity. Since the projected distance from the kilonova to the G1 center is $d_{\rm KN} = 13.7 h$, one finds $P(d_{\rm KN}) = 1.7 \times 10^{-5}$. Consequently, the hypothesis that the progenitor stars were in the disk and that the NS natal kicks did not produce a significant displacement is ruled out by more than $4 \sigma$.

\subsection{Supernova natal kicks: modeling}

The progenitors of BNS systems must undergo two supernovae (SNe), with each event imparting a natal kick to the newborn neutron star \citep{janka_2012}. Depending on the magnitude of these kicks, the binary may be disrupted or survive with modified orbital parameters and systemic velocity \citep{gaspari_2024a}. Observations of pulsars show natal kicks peaking at $200$--$400 \ \mathrm{km \ s^{-1}}$ and extending up to $\sim 1000 \ \mathrm{km \ s^{-1}}$ \citep{hobbs_2005, igoshev_2020, kapil_2023, disberg_2025}. BNS survival under such conditions is possible mainly if the pre-SN orbital separation is small ($\lesssim  10 \, R_\odot$) or if the natal kicks are weak \citep{Kalogera:1998df, Igoshev:2021bxr}. BNS progenitors typically undergo a common envelope phase that reduces their orbital separation \citep{giacobbo_2018, vigna_gomez_2018, kruckow_2018, debatry_2021, broekgaarden_2022, iorio_2023, sgalletta_2023}. 
The second SN usually dominates the dynamics, as the tighter orbit makes its kick more effective. In $\sim 80\%$ of modeled cases (Section~\ref{sec:sevn}) the second kick exceeds the first. Given the short interval between the two SNe ($\lesssim 20$ Myr) compared to the long delay to merger (hundreds of Myr), we consider the second natal kick when modeling BNS systemic motion. For the reasons above, we focus exclusively on the second natal kick when modeling the motion of the system. Henceforth, a single effective kick will be considered.

Due to energy conservation, the binary progenitor system satisfies, at any moment after the kick,  
\begin{equation} \label{eq:energyConservation}
  \frac{1}{2} \mathbf v^2 + \Phi = E \, . 
\end{equation}
In the above,  $E$ is a constant, $\mathbf v = \dot{\mathbf r}$ is the center-of-mass velocity vector of the binary system and $\Phi$ is the total gravitational potential due to the galaxy. Using cylindrical coordinates, the velocity before the kick ($\mathbf v_0)$ and the one after the kick ($\mathbf v$) are decomposed as, 
\begin{align}
    & \mathbf v_{0} = V_{\rm rot}(R_0) \, \hat \theta = \frac{\ell_{z_0}}{R_0} \hat{\theta} \, ,\\
    & \mathbf v =  v_R \hat R + \frac{\ell_z}{R} \hat{\theta} + v_z \hat z,
\end{align}
where $R_0$ is the radial position just before the kick, $\ell_{z_0}$ is the $z$-component of the specific angular momentum ($\boldsymbol \ell$, with $\boldsymbol \ell = \mathbf{r} \times \dot{\mathbf{r}}$) before the kick, and $\ell_z$ is the same quantity after the kick.  After the kick, there are no relevant encounters and the binary system moves freely, thus $\ell_z$ and $E$ are conserved. 

Taking the difference of eq.~\eqref{eq:energyConservation} just after the kick and at a later time, 
\begin{align} \label{eq:vr}
    & v_R^2(R,z) + v^2_z(R,z)   =   \\[.2cm] 
    & \;\;\;\; K_R^2 + K_z^2 +  \left( \frac{1}{R^2_0} - \frac{1}{R^2}\right) \ell_z^2  + 2 \Phi(R_0,0) - 2 \Phi(R,z) \, , \nonumber
\end{align}
where $K_R$ and $K_z$ are components of the kick velocity vector $\mathbf K$. The relation between the component $K_\theta$ and $\ell_z$ reads, $\ell_z = R_0 \left( V_{\rm rot} + K_\theta\right)$.

Regarding the kick direction, for fixed $|\mathbf K|$, we note that:
$i$) $\ell_z$ and the kinetic energy are maximized (minimized) when $\mathbf{K}$ is aligned (anti-aligned) with the rotational velocity $\mathbf{V}_{\rm rot}$. The average case corresponds to $\mathbf K$ aligned with $\hat{\mathbf R}$.
$ii$) From Eq.~\eqref{eq:vr}, for $K_z = 0$, if the binary system reaches large radial distance ($R \gg R_0$), the radial velocity $v_R$ is maximized when $\mathbf{K}$ is aligned with $\mathbf{V}_{\rm rot}$. 
$iii$) Since the host inclination is small ($\sim 34^\circ$), a kick primarily in the $\hat{z}$ direction is particularly unlikely considering the observed angular offset. That would increase the physical displacement by a factor of about two. We will not consider this possibility.

In the $z=0$ plane, the gravitational potential for the dark matter, the stellar and gas contributions are found from
\begin{align}
  & \Phi_{\rm NFW}(R) = - G \frac{4 \pi \rho_s R_s^3}{R} \ln \left(1 + \frac{R}{r_s} \right) \, , \\[.2cm]
  & \Phi_{\rm exp}(R) = - G \pi \Sigma_0 R (I_0 K_1 - I_1 K_0) \, . \label{eq:PhiExpz0}
\end{align}
$\Phi_{\rm exp}$ is the gravitational potential generated by a thin exponential disk with disk-scale length $h$ and central surface density $\Sigma_0$ \citep{BinneyTremaine}. It can be used for both stellar and gas components. The functions $I_n$ and $K_n$ are modified Bessel functions evaluated at $R/(2h)$. The total Newtonian potential reads, $\Phi = \Phi_{\star} + \Phi_{\rm gas} + \Phi_{\rm NFW}$.

\begin{figure*}[ht]
    \begin{center}
        \begin{tikzpicture}
            \node (fig)  {\includegraphics[width=.75\linewidth]{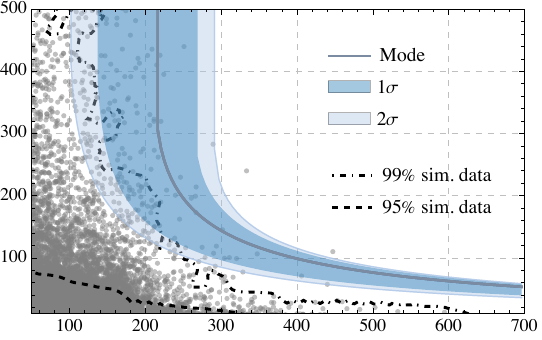} };
            \node[below=of fig,  yshift=1cm, xshift=0.2cm] { \large $K_R$ (km/s)};
            \node[left=of fig, xshift=0.8cm, yshift=1.5cm, rotate=90] {\large $\Delta t$ (Myrs)};
        \end{tikzpicture}
    \end{center}
    \caption{The time interval $ \Delta t $, as a function of the radial kick velocity component $K_R$, for the binary system to travel from an initial radius $ R_0 $ to the observed kilonova location. The solid curve labeled ``mode'' assumes the galaxy parameters ($\delta_\star, \Delta i,  \log c, \log r_{200}$) are given by their best-fit values and that the initial radius is the most probable one, $R_0 = h$.  The shaded regions are the 1$\sigma$ and 2$\sigma$ credible intervals obtained by marginalizing over galaxy parameters and the initial position. The gray dots are simulated events from \sevn\ (several of them have close to zero or negative $K_R$ values and do not appear in the plot). The black dashed and dot-dashed curves show the 95\% and 99\% smooth limits, respectively, obtained from a KDE of the simulation data points.
    } 
    \label{fig:v0XtimeNFW}
\end{figure*}

The time it takes for a particle to travel from $R_0$ to a radial position $R = d_R$  can be found from 
\begin{equation} \label{eq:Deltat}
    \Delta t = \int_{R_0}^{d_R} \frac{dR}{v_R} \, .
\end{equation}

Let $\mathbf d$ be the position vector from the center of host galaxy to the kilonova position. We use $d_z$ and $d_R$ to refer to the distance components in cylindrical coordinates, $d_\perp$ the projected line-of-sight distance, and $d_r = |\mathbf d|$ the distance using the spherical radius $r$. To unveil the relation between $d_R$ and $d_\perp$ we use
\begin{equation}
  d_{\perp}^2 = d_{R}^2 \cos^2(\theta)  + d_R^2 \left( \sin(\theta) \cos(i) - \frac{d_z}{d_R} \sin(i) \right)^2  \, ,  
\end{equation}
where $i$ is the galaxy inclination and $\theta$ is the cylindrical coordinate angle with $\theta=0$ corresponding to the major-axis direction. In the case considered, $i=33.9^\circ$ and $\theta \sim 30^\circ$, implying that $d_{\perp} \sim 0.96 d_{R}$ for $d_z=0$. Since $d_\perp$ is already large, we consider that $d_r$ should not be much larger.  Henceforth, we adopt the simplifying assumption that $d_z\approx0$.

In Fig.~\ref{fig:v0XtimeNFW} we plot the time interval $\Delta t$ between the starting orbit at $R_0$ and the detected position at $R = d_{\perp} = 13.7 h$. 
The solid line, designated in the figure by ``mode'', correspond to the most probable configuration, that is, the galaxy parameters are at their best-fit values, as in Fig.~\ref{fig:nfw_rot_curve}, and the initial radial position is the most probable \eqref{eq:pdfR}, $R_0 = h$. The 1$\sigma$ and 2$\sigma$ HDRs marginalize over the galaxy parameters and $R_0$, with the constraint $R_0 \leq 5 h$. Larger values of $R_0$ have probability of 4\% according to Eq.~\eqref{eq:pdfR}, whereas here we consider only likely initial configurations. 
The same plot also shows regions that are derived from BNS population simulations, which are described in the next subsection.

\subsection{Supernova natal kicks: simulations} \label{sec:sevn}

To compare our results with the expected properties of the of BNS population, we use the archival dataset \textit{SEVNbenchmark2401}~\citep{iorio_2025_16587145}\footnote{Available on Zenodo: \href{https://doi.org/10.5281/zenodo.16587145}{https://doi.org/10.5281/zenodo.16587145}}.
This dataset includes the evolution of $4 \times 10^7$ massive binaries (with primary\footnote{The primary star is the initial more massive star in a binary.} masses greater than 5 M$_\odot$) for 15 representative metallicities in the range $0.0001 \leq Z \leq 0.03$\footnote{$Z$ denotes the mass fraction of elements heavier than helium.}, resulting in a total of $6 \times 10^8$ binaries. The simulations were performed using the state-of-the-art population synthesis code \sevn\footnote{\sevn\ version 2.7.5 (commit \href{https://gitlab.com/sevncodes/sevn/-/commit/7bb74a899cde659b75284aceb3202bc7cc5a6158}{7bb74a89})} \citep{spera_2019, mapelli_2020, iorio_2023}.

For this dataset, we adopted the {\sc Parsec} stellar evolution models presented in \citet{costa_2025}.
\sevn\ models stellar and binary evolution processes (including supernovae, mass transfer, stellar tides, and common envelope evolution) using analytic and semi-analytic prescriptions \citep{iorio_2023}.
In this work, we adopt the delayed supernova explosion mechanism described by \citet{fryer_2012}, while supernova natal kicks are implemented following the prescriptions of \citet{giacobbo_2020}. In this model, the kick velocity is drawn from a Maxwellian distribution with a root-mean-square speed of $265 \ \mathrm{km \ s^{-1}}$ \citep{hobbs_2005}, and subsequently rescaled by a factor proportional to $m_\mathrm{ej}/m_\mathrm{rem}$, where $m_\mathrm{ej}$ is the ejected mass and $m_\mathrm{rem}$ is the remnant mass. The kick direction is randomly sampled assuming spherical symmetry. In a recent analysis, \cite{disberg_2025} revised the kick velocity distribution of isolated neutron stars, finding a slightly lower root-mean-square value of 217 km~s$^{-1}$. However, in our rescaled kick models, surviving double neutron stars populate a low-velocity sub-distribution created by the rescaling itself (see \citealt{giacobbo_2020}), so this difference is not relevant for this work.
For a detailed description of the catalog, we refer to \cite{marinacci_2025}, while  a comprehensive description  of the \sevn \ code can be found in \cite{iorio_2023}.

From the catalog, we extract all the BNS systems, totaling $5.85 \times 10^5$, for all metallicity range and consider their natal kick velocity magnitudes, $K$, along with the corresponding delay times, $\Delta t$, defined as the time between binary formation and merger (e.g., kilonova event).

For the analysis in this work, only the radial component of the kick velocity $K_R$ is relevant. For each simulated effective kick, with absolute value $K_{(i)}$, we associate a random angle $\theta_{(i)} \in (0, \pi)$ between the kick vector $\mathbf{K}$ and the radial direction $\hat{R}$, such that $K_{R,(i)} = K_{(i)} \cos \theta_{(i)}$.
Since the angular dependence of an isotropic distribution satisfies $f(\theta, \phi)  =  \sin(\theta) / 4 \pi$), for each kick, we randomly choose $\cos \theta_{(i)}$ in the interval $(-1,1)$.

Using this procedure, we construct the distribution of the pair $(K_R, \Delta t)$ from simulated data. Each simulated BNS system is thus represented as a point in the $K_R$–$\Delta t$ plane. We identify the highest density regions containing about 95\% and 99\% of the simulated systems using a Gaussian kernel density estimation (KDE), and display the results in Fig.~\ref{fig:v0XtimeNFW}. 
This figure shows an overlap between the two approaches (the $K_R$–$\Delta t$ plane from galaxy-mass modeling and the one from simulations), but it also shows a tension. To quantify it, we count the relative number of simulated events that are within the 2$\sigma$ region of the galaxy-modeling approach and we find 0.1\%. Hence, we conclude that the progenitor binary system may have originated in the disk of the G1 galaxy, although this possibility requires unlikely conditions.

%%%%%%%%%%%%%%%
\section{Conclusions}

GRB\,230307A, if it is indeed associated with a kilonova, represents an extreme case. Its kilonova has a minimum distance of  $\sim$40 kpc (or $\sim$13.7 $h$) from putative host (G1). The host is a rotationally supported disk galaxy with the following properties that we unveil here and are essential for the mass modeling: $i.$ its stellar mass is found to be $\log_{10}(M_\star/M_\odot) = 4.8 \times 10^9$. $ii$. The stellar surface brightness is well described by an exponential disk profile with disk scale length\footnote{This is the main value we use here, from GALFIT. Using {\sc Morfometryka}, we found a smaller value, $h = 1.72''$; but both profiles have a good agreement, apart at close to the galaxy center, Fig.~\ref{fig:SBB}.} $h = 2.19'' \approx 2.90 \mbox{ kpc}$. $iii$. The atomic gas mass (HI + He) is not measured directly, but it is estimated to be $M_{\rm gas} \approx 1.8 \times 10^9$ M$_\odot$. Using JWST data and comparisons with known globular clusters, we show that a globular-cluster origin is unlikely (Sec.~\ref{sec:globular}), favoring a most likely scenario in which the progenitor formed via isolated binary evolution. The large offset, in this case, implies a substantial natal kick ($\gtrsim 200$ km/s) together with a relevant coalescence time ($\Delta t \gtrsim 200$ Myrs): such combination is not commonly realized in the BNS simulations, as the coalescence time and the natal kick magnitude are anti-correlated, as shown in Fig.~\ref{fig:v0XtimeNFW}. We find that only 0.1\% of the simulated BNS systems have radial kick velocity ($K_R$) and coalescence time  compatible (at $2\sigma$-level) with the requirements from the galaxy-mass modeling we did here. 

We find that both a GC origin and a typical disk origin are unlikely, with probabilities of order $\sim 0.1\%$ in both cases. Our analysis indicates that this GRB, if associated with a kilonova from the merger of two neutron stars, did not originate from typical initial conditions: either the neutron stars were born in an atypical globular cluster (given its possible location in the plane of Fig.~\ref{fig:gclf}), or the BNS kick occurred at an atypical galaxy location or with an atypical initial velocity. Considering our kick-velocity direction assumptions, we we have considered negligible azimutal component with respect to the galaxy disk, this since a large component would enlarge the travel distance and hence enlarge the necessary kick magnitude. For the velocity component in the galaxy plane, we studied the average case: neither in the same direction of the galaxy circular velocity, nor opposite to it.

Future multi-wavelength observations of similar events will be essential to map the full distribution of host galaxy properties, offsets, progenitor ages, and to connect these to the underlying population of merging compact objects. GRB\,230307A  case point out that compact binary mergers, while governed by the same fundamental physics, can appear in different galactic contexts.

\acknowledgments
The authors made use of Sci-Mind servers machines developed by the CBPF AI LAB team and would like to thank A. Santos, P. Russano, and M. Portes de Albuquerque for all the support in infrastructure matters.
The authors acknowledge the computational resources made available by the Sci-Com Lab of the Physics Department at UFES, with support from FAPES, CAPES, and CNPq (Brazil).
CRB acknowledges the financial support from CNPq (316072/2021-4) and from FAPERJ (grants 201.456/2022 and 210.330/2022) and the FINEP contract 01.22.0505.00 (ref. 1891/22). This material is based upon work supported by NSF Grant No. 2308193. 
DCR thanks CBPF and \textit{Núcleo de Informação C\&T e Biblioteca} (NIB/CBPF) for hospitality, where part of this work was done. He also acknowledges CNPq, \textit{Fundação de Amparo à Pesquisa e Inovação do Espírito Santo} (FAPES, Brazil) and \textit{Fundação de Apoio ao Desenvolvimento da Computação Científica} (FACC, Brazil) for partial support.  
AC aknowledges the FAPERJ (Brazil) grants E26/202.607/2022 and 210.371/2022(270993) and CNPq. 
DRD acknowledges the financial support from CNPq (grant 313040/2022-2).
GI was supported by a fellowship grant from the \textit{la Caixa} Foundation (ID 100010434). The fellowship code is LCF/BQ/PI24/12040020. 
CDK gratefully acknowledges support from the NSF through AST-2432037, the HST Guest Observer Program through HST-SNAP-17070 and HST-GO-17706, and from JWST Archival Research through JWST-AR-6241 and JWST-AR-5441.
LLN thanks \textit{Funda\c{c}\~ao de Amparo \`a Pesquisa do Estado do Rio de Janeiro} (FAPERJ) for granting the postdoctoral research fellowship E-40/2021(280692).
AEAC thanks D. Krajnović and Muñoz López,C. for the helpful discussions.

The JWST data presented in this article were obtained from the Mikulski Archive for Space Telescopes (MAST) at the Space Telescope Science Institute. The specific observations analyzed can be accessed via \dataset[doi: 10.17909/eadg-ky78]{https://doi.org/10.17909/eadg-ky78}.
\software{
    \texttt{astropy} \citep{astropy:2013,astropy:2018},
    \texttt{BAYESTAR} \citep{bayestar,Singer_2016,Singer_supp},  
    \texttt{eazy-py} \citep{Brammer_2008},
    \texttt{GALFIT} \citep{galfit_peng2002},
    \texttt{IFSCube} \citep{2020zndo...3945237R,2021MNRAS.507...74R},
    \texttt{KDEpy} \citep{kdepy},  
    \texttt{Morfometryka} \citep{morfometryka},
    \texttt{pPXF} \citep{Cappellari2023},
    \texttt{TOPCAT} \citep{topcat}.
}

\newpage % Added \newpage to avoid the acknowledgements to be printed over the references.
\bibliographystyle{yahapj_twoauthor_arxiv_amp}
\bibliography{references}

\restartappendixnumbering 

\appendix

\input{appendixA}

\input{appendixB}
\end{document}

%% file: appendixA.tex
\section{Jwst photometry and the galaxy morphology} \label{app:photometry}

\begin{figure*}[!]
\begin{center}
\includegraphics[width=\linewidth]{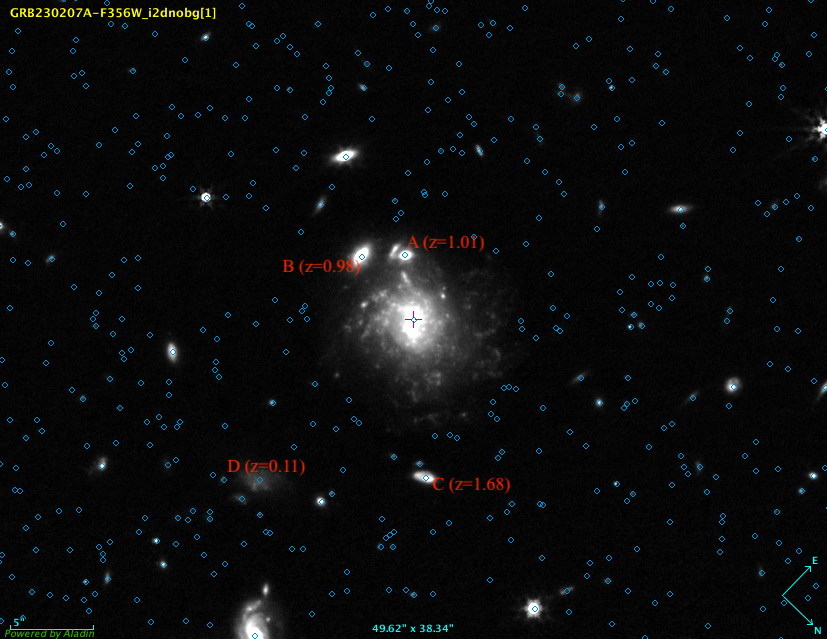}
\end{center}
\caption{$356$ band image of the host galaxy, the blue circles show the objects for which JWST photo-z measurements  are available, see Section \ref{sec:morphology} for more details. Three small galaxies which appear near the  host galaxy ($\le 6"$) have their photo-z written with red labels, and its consistent with $z\ge1$.  A cloudy feature its also present towards south-east with a photo-z consistent with the host galaxy one, given the large error bars, see Fig. \ref{fig:photoz}. 
}
\label{fig:companions}
\end{figure*}

\begin{figure*}[!]
\begin{center}
\includegraphics[width=0.5\linewidth]{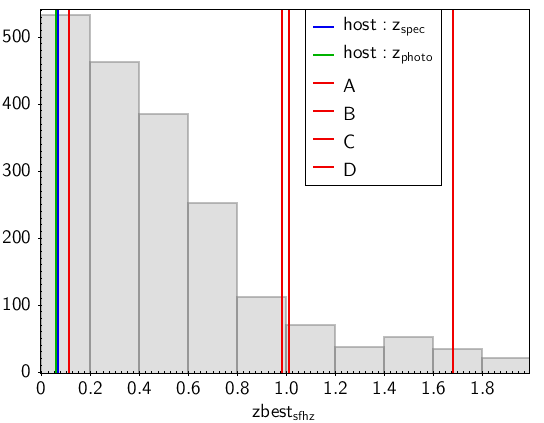}
\end{center}
\caption{Histogram of the photometric redshift retrieved using the JWST photometry within $z\le2$.
Bins of $0.2$ have been chosen to suggest the average size of redshift errors. The green and the blue line show the spectroscopic and photometric redshift of the host galaxy, respectively. The red lines show the redshift of the four neighbouring objects (see Figure \ref{fig:companions} for the exact value of the redshift), only object D is consistent to be associated with the host galaxy, given the redshift errors.}    
\label{fig:photoz}
\end{figure*}

\begin{figure*}[!]
\begin{center}
\includegraphics[width=\linewidth]{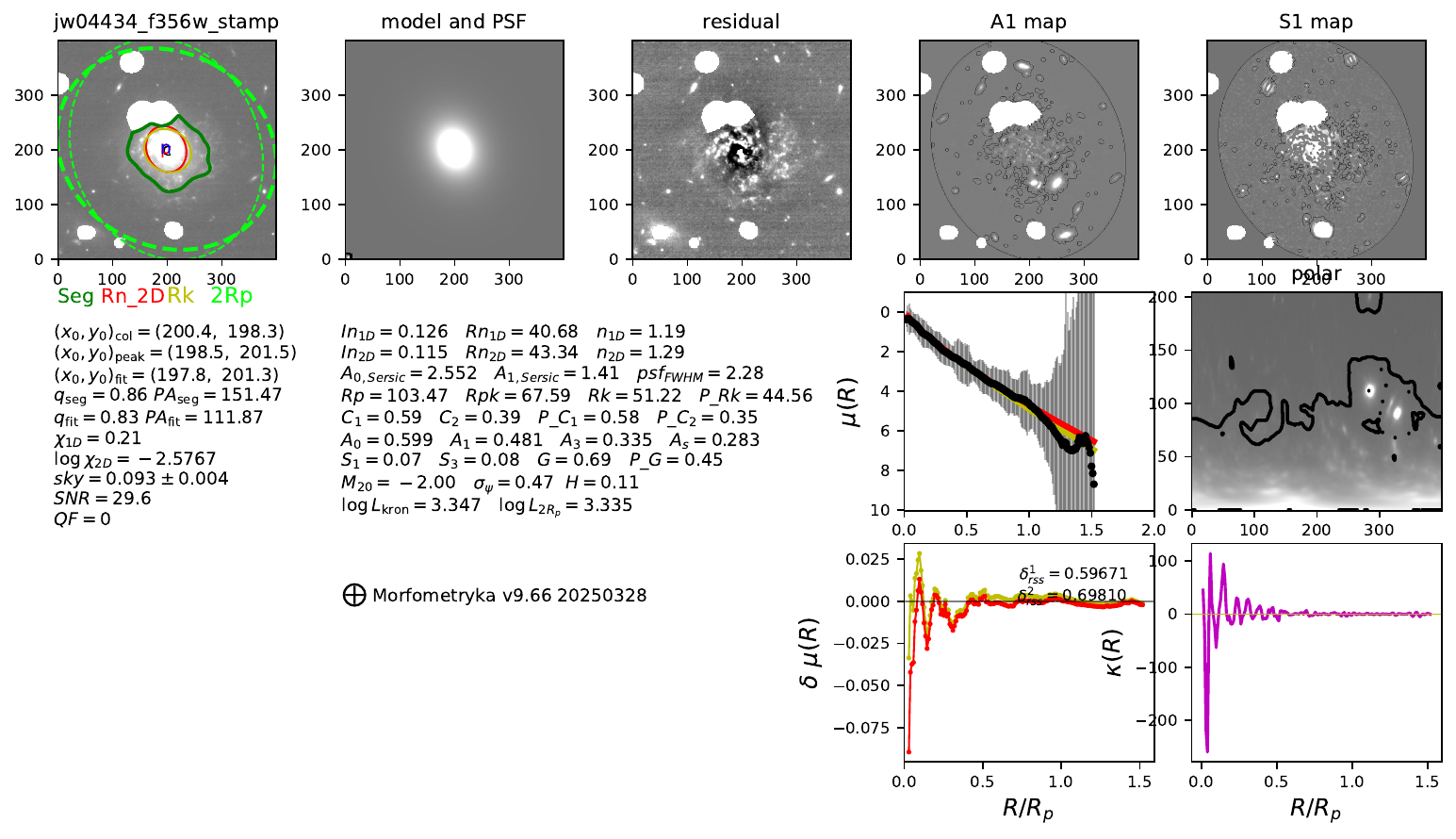}  
\end{center}
\caption{Standard MFMTK output, in the 356 band. The top row shows the galaxy image, first box, and the masked regions (in white). The light green  contour shows the Petrosian region, where the parameters are measured. The recovered parameters are shown below. The second image shows a single Sérsic fit. Both the 1-D and 2-D fit are tabled. The 2-D recovered Sérsic index is $n\simeq 1.3$. In the third box is presented the residual image (image - model), which clearly shows asymmetry in the spiral arms, suggesting it could be a lopsided galaxy. The fourth and fifth panel show the asymmetry map and the smoothness. The other four boxes show the surface brightness profile, the polar map and the second derivative of the galaxy light profile, i.e. the curvature \citep{Lucatelli} that is consistent with a disk galaxy ($k=0$).}
\label{fig:MFMTK}
\end{figure*}

We obtain photometric redshifts for all objects in the field, blue circles in Figure \ref{fig:companions}, using {\tt eazy-py} \citep{Brammer_2008} and the standard set of basis {\tt fsps} templates \citep{conroy2010fsps}. We set a minimum flux error of 5\% to account for residual errors in flux calibration, zero-points and the templates themselves. The catalog of photometric redshift is published as complementary material of this paper, see 
Table \ref{tab:table}.
In Figure \ref{fig:companions} are also marked in red the redshift of three galaxies close to the KN host. The redshift are superimposed to the image. Considering the average uncertainty on the photometric redshift, see Figure \ref{fig:photoz}, neither of these three objects is consistent with being associated with the host galaxy, and have been therefore masked during the galaxy light fitting. MFMTK was used to create the mask. On the other side, object D,  has the same redshift of the host galaxy within errorbars, and given its irregular morphology, could be the remnant of a recent minor merger. The galaxy present a lopsided morphology, with the  arms in the north-east direction being more unwinded, as visible both in Figure \ref{fig:companions} and in Figure \ref{fig:MFMTK}. In the latter, the upper row show the image, the single Sersic model used to constrain the initial conditions of the GALFIT model and the residual, first, second and third panel respectively. The residual image highlights the irregularity of the spiral arms. MFMTK performs a fitting of the 1D and 2D light profile, leaving the Sersic index free, and resulting in a value of $n=1.19$ and $n=1.29$, respectively. These values are slightly higher than $n=1$, used in the modeling of the galaxy potential and imposed in the GALFIT fit. Yet, this increment respect to a pure exponential disk is caused by the central regions of the galaxy, as visible in the last panels of Figure \ref{fig:MFMTK}, which shows the curvature, $\mu$ and its variance $\delta\mu$. The curvature, being the second derivative of the galaxy surface brightness, identifies where the galaxy light changes its curvature, and it is equal to zero for a disk component \citep{Lucatelli}. For $R>0.2R_p$ the galaxy light is well described by a pure exponential, justifying the assumption used in the modeling.

\begin{longrotatetable}    
\vspace*{1cm}
\begin{deluxetable*}{lllllllllllllllll} % deluxtable is suggested here https://journals.aas.org/aastex-v6-3-author-guide/#tables

\tabletypesize{\fontsize{6pt}{7pt}\selectfont}   %font size between scriptsize and tiny
\setlength{\tabcolsep}{3pt}  
\tablecaption{Catalogue of objects detected in the 2 JWST fields covering the Kilonova. The complete table is available online.
The {\tt sfhz} templates were used to determine the photo-z, as they optimize the match between spectroscopic and photometric 
redshift of galaxies in common between MUSE and JWST fields.\label{tab:table}}

\tablehead{
  \colhead{ALPHA\_J2000} &
  \colhead{DELTA\_J2000} &
  \colhead{MAG\_APER} &
  \colhead{MAGERR\_APER} &
  \colhead{KRON\_RADIUS} &
  \colhead{FLUX\_RADIUS} &
  \colhead{FWHM\_IMAGE} &
  \colhead{CLASS\_STAR} &
  \colhead{SNR\_WIN} &
  \colhead{THETA\_IMAGE} &
  \colhead{A\_IMAGE} &
  \colhead{B\_IMAGE} &
  \colhead{FLAGS} &
  \colhead{zbest} &
  \colhead{zbest\_16} &
  \colhead{zbest\_84} &
  \colhead{chi2}
}
\startdata
60.8060 & -75.3803 & 20.3779 & 3.7727E-4 & 4.0000 & 33.8229 & 118.3513 & 0.0286 & 3379.1887 & 88.5551 & 38.1518 & 20.4544 & 3 & 0.4950 & 0.3132 & 0.6649 & 0.0428 \\
60.8024 & -75.3804 & 20.2033 & 4.1870E-4 & 4.0000 & 25.3875 & 39.3311 & 0.0286 & 2746.6938 & -8.4071 & 25.8560 & 18.8615 & 3 & 0.2329 & 0.1186 & 0.3054 & 0.0523 \\
60.8251 & -75.3742 & 21.3246 & 8.4205E-4 & 4.0000 & 8.7975 & 16.7861 & 0.3834 & 1423.0835 & -65.3690 & 12.0840 & 6.8657 & 0 & 1.5306 & 1.2604 & 1.9535 & 5.4611E-5 \\
60.8179 & -75.3755 & 20.6032 & 5.3925E-4 & 4.0000 & 7.9391 & 8.4510 & 0.0329 & 1895.7362 & 6.7433 & 15.5683 & 6.3709 & 0 & 0.8484 & 0.8047 & 1.0461 & 2.0495 \\
60.8310 & -75.3717 & 22.1681 & 1.5731E-3 & 4.0000 & 9.8047 & 19.9286 & 0.0476 & 864.0209 & -43.8627 & 12.0867 & 6.1930 & 0 & 1.2052 & 1.0974 & 1.7343 & 0.0112 \\
60.8402 & -75.3692 & 24.1403 & 8.2513E-3 & 4.0667 & 11.6437 & 27.8955 & 1.3002E-3 & 194.7263 & -59.3981 & 10.9654 & 6.9585 & 2 & 0.3039 & 0.1939 & 0.4588 & 1.7779 \\
60.8133 & -75.3762 & 22.3968 & 1.8833E-3 & 4.0000 & 9.6289 & 21.5412 & 0.0327 & 774.9697 & -76.1240 & 10.7791 & 6.2232 & 0 & 1.3989 & 1.2831 & 2.1356 & 2.7546E-29 \\
60.8207 & -75.3733 & 20.3947 & 4.8719E-4 & 4.0000 & 21.5885 & 112.4668 & 0.8414 & 2503.4436 & 67.5095 & 21.8716 & 16.0458 & 17 & 0.2307 & 0.1501 & 0.2910 & 0.7537 \\
\dots & \dots & \dots &\dots & \dots & \dots & \dots & \dots & \dots & \dots & \dots & \dots & \dots & \dots & \dots & \dots & \dots \\
\enddata

\end{deluxetable*}
\end{longrotatetable}

%% file: appendixB.tex
\section{Rotation curve and mass model derivation} \label{app:rotDerivation}

To find the rotation curve ($V_{\rm rot}$), the line-of-sight velocity ($V_{\rm los}$) is deprojected using the ratio between the major and minor axes found in Sec.~\ref{sec:morphology}, which yields the inclination $i = (0.59 \pm 0.04) \mbox{ rad}$. 
Moreover, the approaching and receding sides need to be combined. To this end, we follow the same procedure of \citet{2009A&A...493..871S} to estimate the central values and corresponding uncertainties. More precisely, we do the following steps: $i$) the kinematical center is set to be at $R=0$, with $R \ge 0$. $ii$) The central values of each $V_{\rm rot}$ data point, denoted by $V_0$, are found from a weighted  mean of the two closest data points (in the $R$ coordinate) from the approaching and receding sides. $iii$) The uncertainties of $V_{\rm rot}$, denoted by $\sigma$, are enlarged due to the rotation curve asymmetry.

\begin{figure*}
\begin{center}
\includegraphics[width=.45\linewidth]{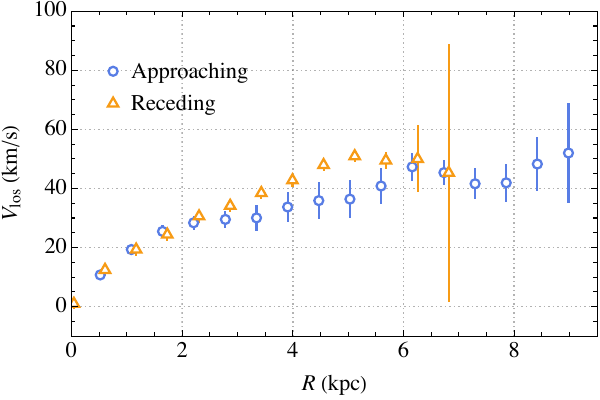} \quad
\includegraphics[width=.45\linewidth]{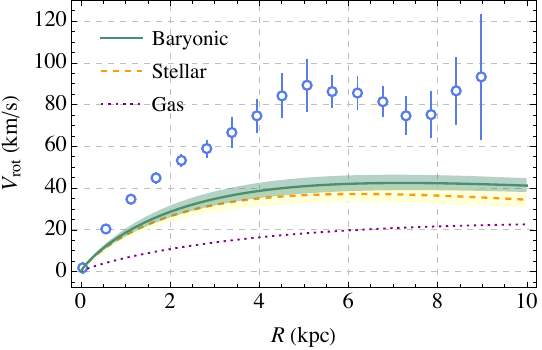}
\end{center}
\caption{\textit{Left plot.} The line-of-sight velocity for the approaching and receding sides with $R=0$ as the kinematical center. \textit{Right plot.} The inclination-corrected mean rotation curve (blue circles with error bars).   The plot also contains the stellar contribution to the circular velocity (dashed orange line, with mass uncertainty indicated in the yellow band), the estimated gas (HI and He) contribution (dotted purple curve) and the total baryonic contribution to the rotation curve (solid green curve) with uncertainty.}
\label{fig:rot_curve}
\end{figure*}

More specifically, following \citet{2009A&A...493..871S}, let each data point of the  approaching and receding velocities be written as $v_a \pm \sigma_a$ and $v_r \pm \sigma_r$. The quantity $\sigma_A$ quantifies the asymmetry as $\sigma_A = \mbox{max} \left ( |v_a - V_0|, |v_r - V_0| \right )$. Then, each $V_{\rm rot}$ data point is given by
\begin{align}
  V_{\rm rot} = \, & V_0 \pm  \sigma \\ 
  = \, & \frac{v_a \, \sigma_a^{-2} + v_r \, \sigma_r^{-2}}{\sigma_a^{-2} + \sigma_r^{-2}} \pm \sqrt{\frac{1}{\sigma_a^{-2} + \sigma_r^{-2} } + \left(\frac{ \sigma_A}{2}\right)^2 } \, , \nonumber
\end{align}
where the approaching and receding velocities read $v_a \pm \sigma_a$ and $v_r \pm \sigma_r$, while $\sigma_A = \mbox{max} \left ( |v_a - V_0|, |v_r - V_0| \right )$. In Table \ref{tab:vrot} we provide the rotation curve here derived, while in Fig.~\ref{fig:rot_curve} we show the approaching and receding data besides the final rotation curve with the baryonic contributions over plotted. In that table and figure we have converted from arcsec to kpc. This conversion was done assuming that the distance to the galaxy is $D = 291$ Mpc at the redshift $z = 0.0647$ \citep{2024Natur.626..742Y}, thus 1 arcsec corresponds to $D \, (1+z)^{-1} \, (180/\pi \times 60 \times 60)^{-1} = $ 1.325 kpc.

\begin{deluxetable}{rrr}
\setlength{\tabcolsep}{15pt}  
\tabletypesize{\small}
\tablecaption{The rotation curve of G1. The first column is the radial position, the second the rotation velocity and the third the corresponding uncertainty.\label{tab:vrot}}
\tablewidth{0pt}
\tablehead{
  \colhead{$R$ (kpc)} &
  \colhead{$V_{\rm rot}$ (km/s)} &
  \colhead{$\sigma$ (km/s)}
}
\startdata
0.04 & 1.7 & 2.5 \\
 0.56 & 20.3 & 2.0 \\
 1.13 & 34.6 & 1.9 \\
 1.69 & 44.8 & 2.3 \\
 2.26 & 53.2 & 2.7 \\
 2.82 & 59. & 4. \\
 3.39 & 67. & 7. \\
 3.95 & 75. & 8. \\
 4.51 & 84. & 10. \\
 5.08 & 89. & 12. \\
 5.64 & 86. & 8. \\
 6.21 & 85. & 8. \\
 6.77 & 81. & 7. \\
 7.29 & 75. & 9. \\
 7.86 & 75. & 11. \\
 8.42 & 87. & 16. \\
 8.98 & 93. & 30. \\
\enddata
\end{deluxetable}

We proceed with Bayesian inference using a grid. This procedure is robust and transparent, without raising issues about convergence. The disadvantage of using the grid method, in comparison to MCMC, is its large computational time when applied to problems with many parameters. Since here we deal with four parameters, the computational time is not a relevant issue. Its final precision is given by the grid spacing, which here we adopt 0.01 for all parameters. In the following, we present the MAP estimate as the central value, together with the corresponding $1\sigma$ and $2\sigma$ uncertainties from the marginalized posteriors for each parameter,
\begin{align}
&\delta_\star = 1.10^{+0.22\,+0.28}_{-0.19\,-0.35} \, ,\nonumber \\[.1cm]
& \Delta i \; [\mbox{rad}] = 0.01^{+ 0.03 \, + 0.07}_{-0.03 \, - 0.06} \, ,\nonumber \\[.1cm]
&\log c = 0.75^{+0.12\,+0.22}_{-0.12\,-0.26} \, ,\label{eq:NFWparameterResults} \\[.1cm]
&\log(r_{200}) \; \mbox{[kpc]} = 2.64^{+0.18\,+0.36}_{-0.16\,-0.30} \, .\, .\nonumber
\end{align}
All the intervals above are the highest density probability intervals.

The (MAP) best fit and the marginalized posteriors can be found in Fig.~\ref{fig:v0XtimeNFW}.

\section{Analysis of Stellar Ages and Metal Abundances}

To derive the stellar population parameters of the galaxy, we used MUSE observations of the GI galaxy, following the methodology described in \citet{lopez2025multiple}. The spectrum was extracted within one effective radius (Reff = 3.67$^{\prime\prime}$), as estimated from JWST imaging, with a signal-to-noise ratio of 9. 

The spectral fitting was performed using the Penalized PiXel-Fitting algorithm (\textsc{pPXF}\footnote{\url{https://pypi.org/project/ppxf/}}), which derives stellar population properties by fitting a linear combination of simple stellar population templates to the observed spectrum \citep{Cappellari2023}. We adopted the E-MILES stellar library \citep{vazdekis2016uv}, which is based on Padova isochrones and assumes a Salpeter initial mass function (IMF). This library was chosen due to its moderately high spectral resolution (FWHM = 2.5~\AA) over the 3540–7500~\AA\ range, which closely matches that of the MUSE data.

Before fitting each galaxy, we restricted the template set to stellar populations younger than the age of the Universe at the galaxy’s redshift, ensuring that the oldest components used in the fit are physically consistent with the cosmic timeline. To minimize contamination from strong OH sky emission lines, we restricted the spectrum to the rest-frame wavelength range between 4350 and 6850~\AA.
To focus on the stellar continuum, we masked all emission lines and excluded outliers with deviations greater than 3$\sigma$ in relative error within the spectral range.  In addition, we used a 4th-order multiplicative polynomial and no additive polynomial, in order to adjust the continuum shape without altering the absorption line strengths.

To estimate uncertainties on the derived stellar population parameters we used the bootstrap approach \citep{davidson2008wild}. We first performed an unregularized fit to obtain the best-fit model and residuals. Then, we generated N= 500 synthetic spectra by perturbing the model spectrum with noise drawn randomly from the residual distribution, and fitted each one with pPXF (again with zero regularization). This specific number of realizations ($N = 500$) was chosen following \citet{lopez2025multiple} after testing several values and confirming that the derived statistics had converged. The final stellar age and total metallicity of each galaxy were calculated as the average over the 500 bootstrap iterations. For each realization $j$, the light-weighted stellar age and metallicity were computed using the weights assigned to each simple stellar population (SSP) model. The final values are given by:

\begin{equation}
\langle \log_{10}(\mathrm{Age}) \rangle = \frac{1}{N} \sum_{j=1}^{N} \left[ \frac{ \sum_{i=1}^{n} \log_{10}(\mathrm{Age}_i) \, w_i }{ \sum_{i=1}^{n} w_i } \right]_j
\end{equation}

\begin{equation}
\langle \mathrm{[M/H]} \rangle = \frac{1}{N} \sum_{j=1}^{N} \left[ \frac{ \sum_{i=1}^{n} \mathrm{[M/H]}_i \, w_i }{ \sum_{i=1}^{n} w_i } \right]_j
\end{equation}

where $N$ is the total number of simulated spectra, $n$ is the number of SSP templates used in each fit, $\mathrm{Age}_i$, $[\mathrm{M/H}]_i$ and $w_i$ are the age, metallicity and weight of the $i$-th SSP model, respectively. The associated uncertainties were derived as the standard deviation of the 500 bootstrap realizations.

\begin{figure}
        \centering
        \includegraphics[width=0.8\linewidth]{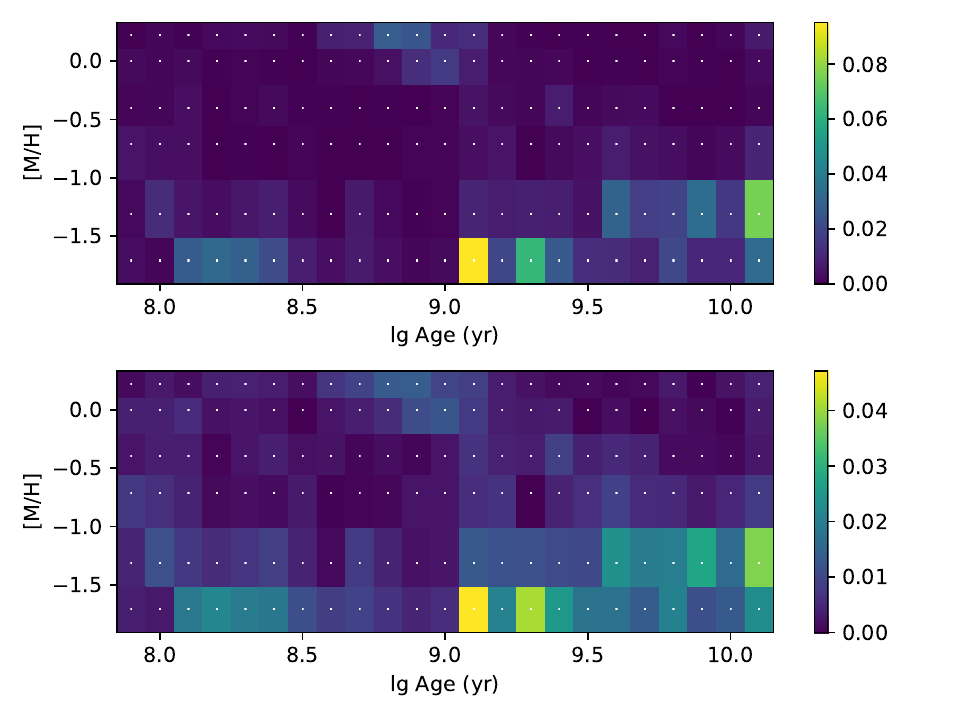}
    \caption{Top: Distribution of the stellar pPXF weights, colored by the light-weighted fraction of different stellar populations with given age and metallicity. Bottom: Standard deviation of pPXF weights after 500 iterations.}
    \label{fig:weights}
\end{figure}

\begin{figure}
    \includegraphics[width=0.49\linewidth]{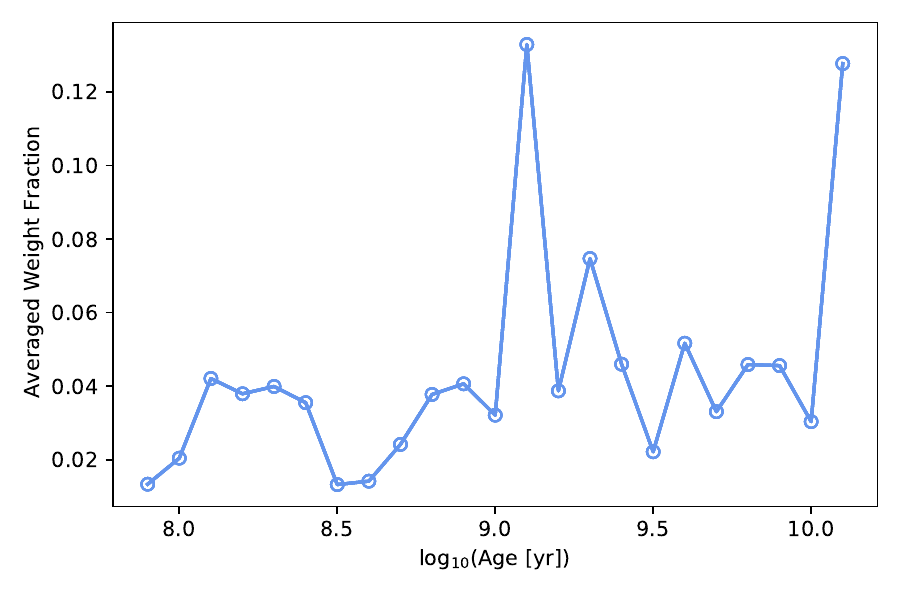}
    \includegraphics[width=0.49\linewidth]{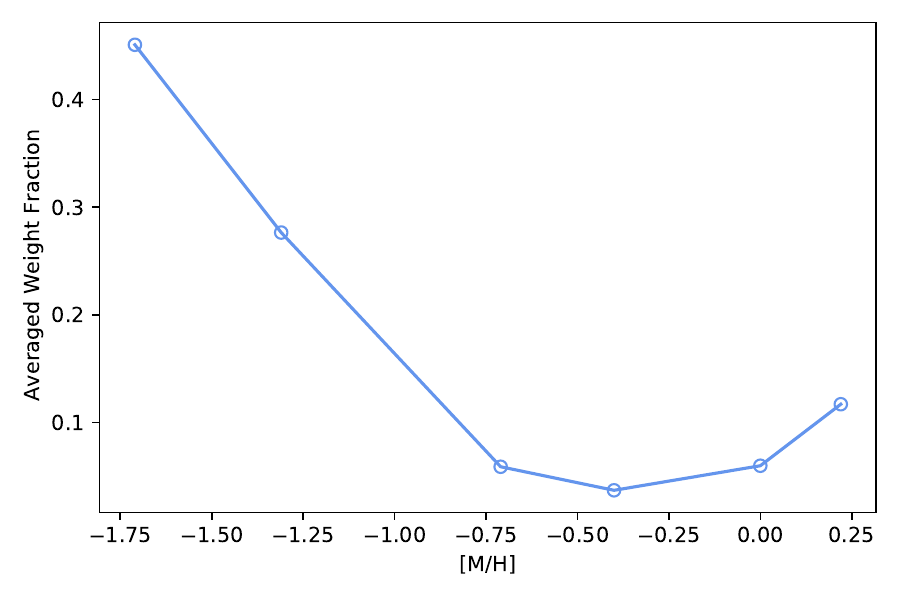}
    \caption{One-dimensional light-weighted distributions of stellar ages (left) and stellar metallicities (right), obtained from spectral fitting using pPXF.}
    \label{fig:stellarpop_metallicity}
\end{figure}

In Figure~\ref{fig:weights}, we show the stellar population weight distribution of the galaxy, averaged over 500 simulated spectra. The results indicate an average light-weighted stellar age of $3.7 \pm 1.3$~Gyr and an average light-weighted metallicity of $[\mathrm{M}/\mathrm{H}] = - 1.03 \pm 0.01$. Moreover, our analysis reveals that the galaxy is primarily composed of three distinct stellar populations. The oldest population, with an age of approximately 10 Gyr, exhibits low metallicity ($[M/H] = 0.001$). A secondary population, roughly 1 Gyr old, has solar metallicity, while a younger population of about 100 Myr displays higher metallicity, around $[M/H] = 0.025$, see Figure \ref{fig:stellarpop_metallicity}.

\section{Stellar Mass Assembly and Star Formation History}

\begin{figure} [h!]
    \centering
    \includegraphics[width=1.0\linewidth]{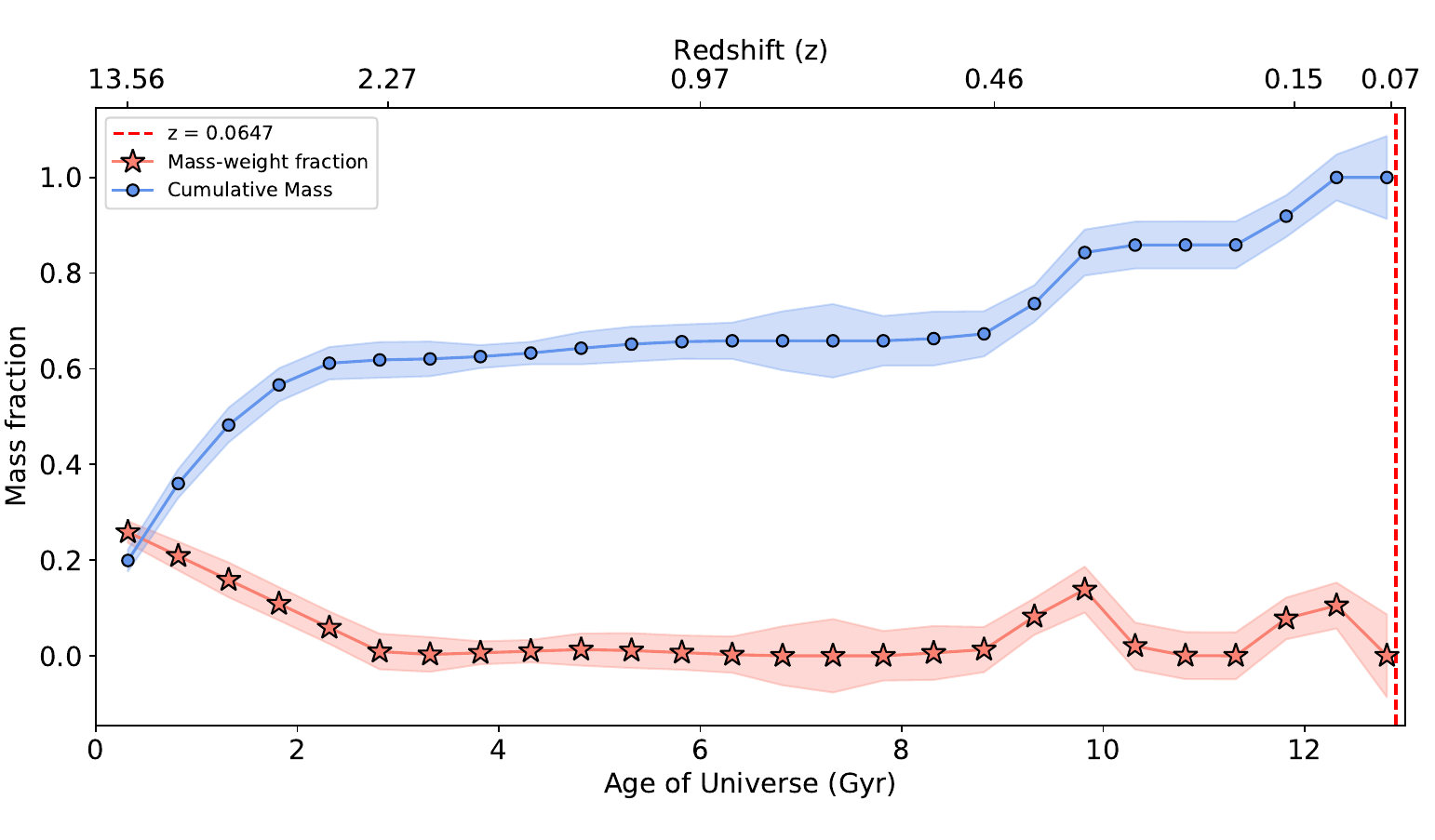}
    \caption{Star formation history (SFH; salmon stars) and cumulative stellar mass (blue line with dot markers). The SFH was smoothed and interpolated in 0.5-Gyr bins to reduce artifacts. The assembled mass fraction relative to galaxy's total mass is shown in the y-axis. The red dashed line indicates the system’s redshift.
    }
    \label{sfh}
\end{figure}

We reconstructed the galaxy’s star formation history (SFH) by combining the average weights of stellar populations derived from spectral fitting. Each normalized weight represents the V-band light-weighted fractional contribution of a simple stellar population (SSP) characterized by a specific age and metallicity. By summing these weights over metallicity at fixed age, we obtained the age distribution and the corresponding the stellar mass fraction, relative to the total stellar mass, of the stellar populations of the galaxy. This approach allows us to trace when in time the different stellar populations formed and how fast the process occurred (see Figure \ref{sfh}). Additionally we integrate the SFH to derive the galaxy's cumulative mass distribution, reveling how the galaxy assembled its stellar mass over cosmic time.  Our findings indicate that the bulk of the galaxy's stellar mass was formed early in the Universe’s history, roughly 10 Gyr ago. The SFH also shows that the galaxy experienced additional mass growth a few Myr ago, which may correspond to the formation of a younger stellar component. However, the current stellar population models do not allow us to distinguish between in situ formation and ex situ accretion scenarios for these recently formed stars.

As highlighted by \citet{lopez2025multiple}, the stellar templates used for spectral fitting are unevenly spaced in age, with a higher density of templates at younger ages. This uneven sampling leads to small \(\Delta t_i\) values in recent epochs, which can artificially inflate the computed SFR, causing spurious peaks. To address this, we interpolate both the SFH and cumulative stellar mass into regular 0.5 Gyr bins. This smoothing procedure provides a more physically meaningful star formation history by mitigating artifacts arising from the variable temporal resolution of the stellar templates. Our results in Figure \ref{sfh} reveal that the galaxy experienced a major burst of star formation early in the Universe’s history, approximately 10 Gyr ago, followed by secondary bursts around 4 and 2 Gyr ago. This later events contributed to the formation of some of the younger stellar populations.

\section{Stellar Mass Estimation} \label{sec:StellarMass}

To derive the galaxy's stellar mass, we combined spectroscopic data from MUSE with photometric measurements from six JWST bands, increasing the overall signal-to-noise ratio to approximately 10. This was achieved using the latest version of the \texttt{pPXF} code \citep{Cappellari2023}. The fitting configuration followed the same setup described previously, but in this case, we simultaneously fit both the stellar continuum and emission lines. Outliers in the spectra were excluded during the fitting process.

Since the photometric and spectroscopic data were not extracted from the same aperture, we scaled the photometry to match the spectral flux. This scaling was performed by computing synthetic photometry from the best-fit spectrum in the JWST filters that overlap with the spectral range, and matching these synthetic fluxes to the observed photometry, as described in \citet{Cappellari2023}.

Once the best-fit spectral weights were obtained, we converted the light-weight contributions into stellar mass by correcting for the intrinsic luminosity of each stellar population. This process included multiplying by the mass fraction in stars and stellar remnants, applying the luminosity distance of the galaxy, and correcting for redshift effects and solar luminosity units. Finally, we applied the photometric scaling factor used to match the flux levels between the spectroscopy and photometry, resulting in a physically consistent estimate of the galaxy’s stellar mass.

We estimated a total stellar mass of $\log_{10}(M_\star/M_\odot) = 9.68 \pm 0.10$. This value differs from that reported by \citet{2024Natur.626..737L}, who found $\log_{10}(M_\star/M_\odot) = 9.37^{+0.04}_{-0.07}$. This discrepancy may arise from differences in the modeling approach: while \citet{2024Natur.626..737L} adopted stellar population models based on the Flexible Stellar Population Synthesis (FSPS) package with a Chabrier IMF, our analysis uses the E-MILES stellar population models assuming a Salpeter IMF.

\begin{figure}
    \centering
    \includegraphics[width=0.9\linewidth]{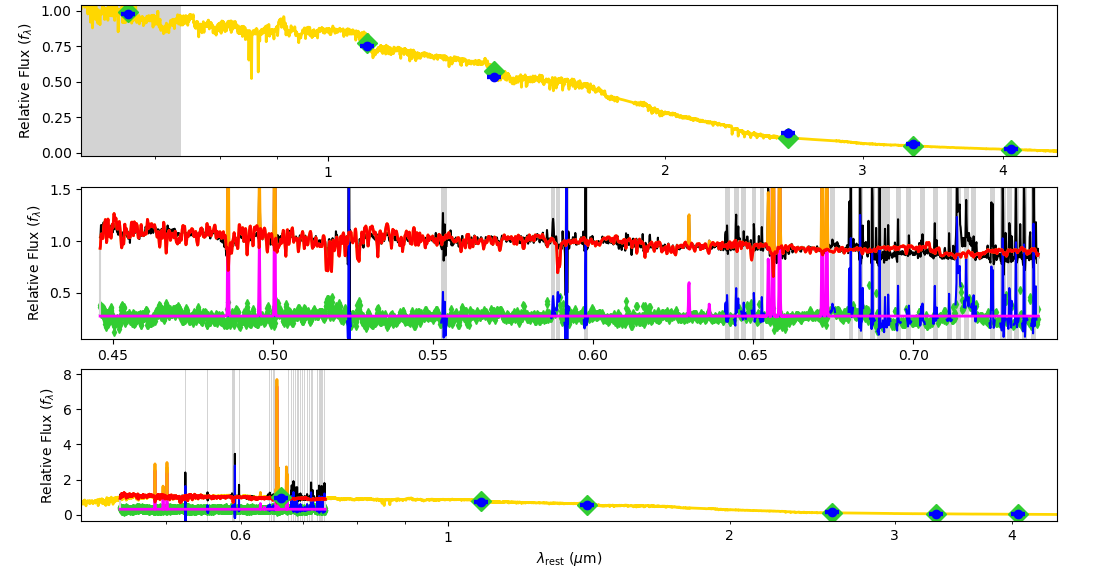}
    \caption{Fit used to estimate the spectral mass of the host galaxy. 
Top: Best-fitting template using E-MILES templates shown as a golden line. Blue points with error bars indicate photometric measurements from JWST. The grey area corresponds to the wavelength range covered by the MUSE spectrum shown in the middle panel. Middle: Observed spectrum in black, pPXF best-fitting total spectrum in orange, emission lines in magenta, masked outliers in blue, and residuals indicated with green diamonds. Bottom: Combined spectra and photometric measurements.
}

    \label{fig:enter-label}
\end{figure}

%% file: main.bbl
\begin{thebibliography}{}
\providecommand\natexlab[1]{#1}
\providecommand\JournalTitle[1]{#1}
\providecommand{\eprint}[1][]{\url{#1}}

\bibitem[{Adams {et~al.}(2024)Adams \& Conselice {et~al.}}]{adams2024epochs}
Adams, N.~J., Conselice, C.~J., Austin, D., {et~al.} 2024, \JournalTitle{The Astrophysical Journal}, 965, 169

\bibitem[{{Astropy Collaboration} {et~al.}(2013){Astropy Collaboration} \& {Robitaille} {et~al.}}]{astropy:2013}
{Astropy Collaboration}, {Robitaille}, T.~P., {Tollerud}, E.~J., {et~al.} 2013, \href{http://dx.doi.org/10.1051/0004-6361/201322068}{\JournalTitle{\aap}, 558, A33}

\bibitem[{{Astropy Collaboration} {et~al.}(2018){Astropy Collaboration} \& {Price-Whelan} {et~al.}}]{astropy:2018}
{Astropy Collaboration}, {Price-Whelan}, A.~M., {Sip{\H{o}}cz}, B.~M., {et~al.} 2018, \href{http://dx.doi.org/10.3847/1538-3881/aabc4f}{\JournalTitle{\aj}, 156, 123}

\bibitem[{{Bertin}(2013)}]{bertin_psfex}
{Bertin}, E. 2013, {PSFEx: Point Spread Function Extractor}, Astrophysics Source Code Library, record ascl:1301.001

\bibitem[{{Bertin} \& {Arnouts}(1996)}]{sextractor}
{Bertin}, E. \& {Arnouts}, S. 1996, \href{http://dx.doi.org/10.1051/aas:1996164}{\JournalTitle{Astronomy and Astrophysics Supplement}, 117, 393}

\bibitem[{Bertola {et~al.}(1991)Bertola \& Bettoni \& Danziger \& Sadler \& Sparke \& de~Zeeuw}]{bertola_testing_1991}
Bertola, F., Bettoni, D., Danziger, J., {et~al.} 1991, \href{http://dx.doi.org/10.1086/170058}{\JournalTitle{Astrophys. J.}, 373, 369}

\bibitem[{Binney \& Tremaine(1988)}]{BinneyTremaine}
Binney, J. \& Tremaine, S. 1988, {Galactic Dynamics (Princeton Series in Astrophysics)} (Princeton University Press)

\bibitem[{{Blakeslee} {et~al.}(2012){Blakeslee} \& {Cho} \& {Peng} \& {Ferrarese} \& {Jord{\'a}n} \& {Martel}}]{Blakeslee12}
{Blakeslee}, J.~P., {Cho}, H., {Peng}, E.~W., {et~al.} 2012, \href{http://dx.doi.org/10.1088/0004-637X/746/1/88}{\JournalTitle{\apj}, 746, 88}, \eprint arXiv:{1201.1031}

\bibitem[{Brammer {et~al.}(2008)Brammer \& van Dokkum \& Coppi}]{Brammer_2008}
Brammer, G.~B., van Dokkum, P.~G., \& Coppi, P. 2008, \href{http://dx.doi.org/10.1086/591786}{\JournalTitle{The Astrophysical Journal}, 686, 1503}

\bibitem[{{Broekgaarden} {et~al.}(2022){Broekgaarden} \& {Berger} {et~al.}}]{broekgaarden_2022}
{Broekgaarden}, F.~S., {Berger}, E., {Stevenson}, S., {et~al.} 2022, \href{http://dx.doi.org/10.1093/mnras/stac1677}{\JournalTitle{\mnras}, 516, 5737}, \eprint arXiv:{2112.05763}

\bibitem[{{Cappellari}(2023)}]{Cappellari2023}
{Cappellari}, M. 2023, \href{http://dx.doi.org/10.1093/mnras/stad2597}{\JournalTitle{MNRAS}, 526, 3273}, \eprint{2208.14974}

\bibitem[{{Chattopadhyay} {et~al.}(2021){Chattopadhyay} \& {Stevenson} \& {Hurley} \& {Bailes} \& {Broekgaarden}}]{debatry_2021}
{Chattopadhyay}, D., {Stevenson}, S., {Hurley}, J.~R., {Bailes}, M., \& {Broekgaarden}, F. 2021, \href{http://dx.doi.org/10.1093/mnras/stab973}{\JournalTitle{\mnras}, 504, 3682}, \eprint arXiv:{2011.13503}

\bibitem[{Conroy \& Gunn(2010)}]{conroy2010fsps}
Conroy, C. \& Gunn, J.~E. 2010, \JournalTitle{Astrophysics Source Code Library}, ascl

\bibitem[{{Conselice} {et~al.}(2020){Conselice} \& {Bhatawdekar} \& {Palmese} \& {Hartley}}]{Conselice_20}
{Conselice}, C.~J., {Bhatawdekar}, R., {Palmese}, A., \& {Hartley}, W.~G. 2020, \JournalTitle{\apj}, 890, arXiv:1907.05361, \eprint arXiv:{1907.05361}

\bibitem[{Conselice {et~al.}(2025)Conselice \& Adams {et~al.}}]{conselice2025epochs}
Conselice, C.~J., Adams, N., Harvey, T., {et~al.} 2025, \JournalTitle{The Astrophysical Journal}, 983, 30

\bibitem[{{Costa} {et~al.}(2025){Costa} \& {Shepherd} {et~al.}}]{costa_2025}
{Costa}, G., {Shepherd}, K.~G., {Bressan}, A., {et~al.} 2025, \href{http://dx.doi.org/10.1051/0004-6361/202452573}{\JournalTitle{\aap}, 694, A193}, \eprint arXiv:{2501.12917}

\bibitem[{{Dalessi} \& {Fermi GBM Team}(2023)}]{2023GCN.33407....1D}
{Dalessi}, S. \& {Fermi GBM Team}. 2023, \JournalTitle{GRB Coordinates Network}, 33407, 1

\bibitem[{Davidson \& Flachaire(2008)}]{davidson2008wild}
Davidson, R. \& Flachaire, E. 2008, \JournalTitle{Journal of Econometrics}, 146, 162

\bibitem[{{Disberg} \& {Mandel}(2025)}]{disberg_2025}
{Disberg}, P. \& {Mandel}, I. 2025, \href{http://dx.doi.org/10.48550/arXiv.2505.22102}{\JournalTitle{arXiv e-prints}, arXiv:2505.22102}, \eprint arXiv:{2505.22102}

\bibitem[{{Ferrari} {et~al.}(2015){Ferrari} \& {de Carvalho} \& {Trevisan}}]{morfometryka}
{Ferrari}, F., {de Carvalho}, R.~R., \& {Trevisan}, M. 2015, \href{http://dx.doi.org/10.1088/0004-637X/814/1/55}{\JournalTitle{\apj}, 814, 55}, \eprint arXiv:{1509.05430}

\bibitem[{Freedman {et~al.}(2019)Freedman \& Madore {et~al.}}]{Freedman_2019}
Freedman, W.~L., Madore, B.~F., Hatt, D., {et~al.} 2019, \href{http://dx.doi.org/10.3847/1538-4357/ab2f73}{\JournalTitle{\apj}, 882, 34}

\bibitem[{{Fryer} {et~al.}(2012){Fryer} \& {Belczynski} \& {Wiktorowicz} \& {Dominik} \& {Kalogera} \& {Holz}}]{fryer_2012}
{Fryer}, C.~L., {Belczynski}, K., {Wiktorowicz}, G., {et~al.} 2012, \href{http://dx.doi.org/10.1088/0004-637X/749/1/91}{\JournalTitle{\apj}, 749, 91}, \eprint arXiv:{1110.1726}

\bibitem[{{Gaia Collaboration} {et~al.}(2023){Gaia Collaboration} \& {Vallenari} {et~al.}}]{GAIADR3}
{Gaia Collaboration}, {Vallenari}, A., {Brown}, A.~G.~A., {et~al.} 2023, \href{http://dx.doi.org/10.1051/0004-6361/202243940}{\JournalTitle{\aap}, 674, A1}, \eprint arXiv:{2208.00211}

\bibitem[{{Gaspari} {et~al.}(2024){Gaspari} \& {Levan} \& {Chrimes} \& {Nelemans}}]{gaspari_2024a}
{Gaspari}, N., {Levan}, A.~J., {Chrimes}, A.~A., \& {Nelemans}, G. 2024, \href{http://dx.doi.org/10.1093/mnras/stad3259}{\JournalTitle{\mnras}, 527, 1101}, \eprint arXiv:{2310.14773}

\bibitem[{{Giacobbo} \& {Mapelli}(2018)}]{giacobbo_2018}
{Giacobbo}, N. \& {Mapelli}, M. 2018, \href{http://dx.doi.org/10.1093/mnras/sty1999}{\JournalTitle{\mnras}, 480, 2011}, \eprint arXiv:{1806.00001}

\bibitem[{{Giacobbo} \& {Mapelli}(2020)}]{giacobbo_2020}
{Giacobbo}, N. \& {Mapelli}, M. 2020, \href{http://dx.doi.org/10.3847/1538-4357/ab7335}{\JournalTitle{\apj}, 891, 141}, \eprint arXiv:{1909.06385}

\bibitem[{{Harvey} {et~al.}(2025){Harvey} \& {Conselice} {et~al.}}]{Harvey2024}
{Harvey}, T., {Conselice}, C.~J., {Adams}, N.~J., {et~al.} 2025, \href{http://dx.doi.org/10.3847/1538-4357/ad8c29}{\JournalTitle{\apj}, 978, 89}, \eprint arXiv:{2403.03908}

\bibitem[{{Hobbs} {et~al.}(2005){Hobbs} \& {Lorimer} \& {Lyne} \& {Kramer}}]{hobbs_2005}
{Hobbs}, G., {Lorimer}, D.~R., {Lyne}, A.~G., \& {Kramer}, M. 2005, \href{http://dx.doi.org/10.1111/j.1365-2966.2005.09087.x}{\JournalTitle{\mnras}, 360, 974}, \eprint arXiv:{astro-ph/0504584}

\bibitem[{{Igoshev}(2020)}]{igoshev_2020}
{Igoshev}, A.~P. 2020, \href{http://dx.doi.org/10.1093/mnras/staa958}{\JournalTitle{\mnras}, 494, 3663}, \eprint arXiv:{2002.01367}

\bibitem[{Igoshev {et~al.}(2021)Igoshev \& Chruslinska \& Dorozsmai \& Toonen}]{Igoshev:2021bxr}
Igoshev, A.~P., Chruslinska, M., Dorozsmai, A., \& Toonen, S. 2021, \href{http://dx.doi.org/10.1093/mnras/stab2734}{\JournalTitle{Mon. Not. Roy. Astron. Soc.}, 508, 3345}, \eprint arXiv:{2109.10362}

\bibitem[{Iorio(2025)}]{iorio_2025_16587145}
Iorio, G. 2025, SEVNbenchmark2401

\bibitem[{{Iorio} {et~al.}(2023){Iorio} \& {Mapelli} {et~al.}}]{iorio_2023}
{Iorio}, G., {Mapelli}, M., {Costa}, G., {et~al.} 2023, \href{http://dx.doi.org/10.1093/mnras/stad1630}{\JournalTitle{\mnras}, 524, 426}, \eprint arXiv:{2211.11774}

\bibitem[{{Ivanova} {et~al.}(2008){Ivanova} \& {Heinke} \& {Rasio} \& {Belczynski} \& {Fregeau}}]{Ivanova08}
{Ivanova}, N., {Heinke}, C.~O., {Rasio}, F.~A., {Belczynski}, K., \& {Fregeau}, J.~M. 2008, \href{http://dx.doi.org/10.1111/j.1365-2966.2008.13064.x}{\JournalTitle{\mnras}, 386, 553}, \eprint arXiv:{0706.4096}

\bibitem[{{Janka}(2012)}]{janka_2012}
{Janka}, H.-T. 2012, \href{http://dx.doi.org/10.1146/annurev-nucl-102711-094901}{\JournalTitle{Annual Review of Nuclear and Particle Science}, 62, 407}, \eprint arXiv:{1206.2503}

\bibitem[{{Jord{\'a}n} {et~al.}(2007){Jord{\'a}n} \& {McLaughlin} {et~al.}}]{Jordan07}
{Jord{\'a}n}, A., {McLaughlin}, D.~E., {C{\^o}t{\'e}}, P., {et~al.} 2007, \href{http://dx.doi.org/10.1086/516840}{\JournalTitle{\apjs}, 171, 101}, \eprint arXiv:{astro-ph/0702496}

\bibitem[{Kalogera {et~al.}(1998)Kalogera \& Kolb \& King}]{Kalogera:1998df}
Kalogera, V., Kolb, U., \& King, A.~R. 1998, \href{http://dx.doi.org/10.1086/306119}{\JournalTitle{Astrophys. J.}, 504, 967}, \eprint arXiv:{astro-ph/9803288}

\bibitem[{{Kapil} {et~al.}(2023){Kapil} \& {Mandel} \& {Berti} \& {M{\"u}ller}}]{kapil_2023}
{Kapil}, V., {Mandel}, I., {Berti}, E., \& {M{\"u}ller}, B. 2023, \href{http://dx.doi.org/10.1093/mnras/stad019}{\JournalTitle{\mnras}, 519, 5893}, \eprint arXiv:{2209.09252}

\bibitem[{{Kilpatrick} {et~al.}(2022){Kilpatrick} \& {Fong} {et~al.}}]{charle_gw170817}
{Kilpatrick}, C.~D., {Fong}, W.-f., {Blanchard}, P.~K., {et~al.} 2022, \href{http://dx.doi.org/10.3847/1538-4357/ac3e59}{\JournalTitle{\apj}, 926, 49}, \eprint arXiv:{2109.06211}

\bibitem[{{Kruckow} {et~al.}(2018){Kruckow} \& {Tauris} \& {Langer} \& {Kramer} \& {Izzard}}]{kruckow_2018}
{Kruckow}, M.~U., {Tauris}, T.~M., {Langer}, N., {Kramer}, M., \& {Izzard}, R.~G. 2018, \href{http://dx.doi.org/10.1093/mnras/sty2190}{\JournalTitle{\mnras}, 481, 1908}, \eprint arXiv:{1801.05433}

\bibitem[{{Lee} {et~al.}(2018){Lee} \& {Kang} \& {Im}}]{Lee18}
{Lee}, M.~G., {Kang}, J., \& {Im}, M. 2018, \href{http://dx.doi.org/10.3847/2041-8213/aac2e9}{\JournalTitle{\apjl}, 859, L6}, \eprint arXiv:{1805.01127}

\bibitem[{{Lelli} {et~al.}(2016){Lelli} \& {McGaugh} \& {Schombert}}]{2016AJ....152..157L}
{Lelli}, F., {McGaugh}, S.~S., \& {Schombert}, J.~M. 2016, \href{http://dx.doi.org/10.3847/0004-6256/152/6/157}{\JournalTitle{\aj}, 152, 157}, \eprint arXiv:{1606.09251}

\bibitem[{{Levan} {et~al.}(2024){Levan} \& {Gompertz} {et~al.}}]{2024Natur.626..737L}
{Levan}, A.~J., {Gompertz}, B.~P., {Salafia}, O.~S., {et~al.} 2024, \href{http://dx.doi.org/10.1038/s41586-023-06759-1}{\JournalTitle{\nat}, 626, 737}, \eprint arXiv:{2307.02098}

\bibitem[{{Lim} {et~al.}(2024){Lim} \& {Peng} {et~al.}}]{Lim24}
{Lim}, S., {Peng}, E.~W., {C{\^o}t{\'e}}, P., {et~al.} 2024, \href{http://dx.doi.org/10.3847/1538-4357/ad3444}{\JournalTitle{\apj}, 966, 168}, \eprint arXiv:{2403.09926}

\bibitem[{L{\'o}pez {et~al.}(2025)L{\'o}pez \& Krajnovi{\'c} {et~al.}}]{lopez2025multiple}
L{\'o}pez, C.~M., Krajnovi{\'c}, D., Epinat, B., {et~al.} 2025, \JournalTitle{Astronomy \& Astrophysics}, 702, A146

\bibitem[{{Lucatelli} \& {Ferrari}(2019)}]{Lucatelli}
{Lucatelli}, G. \& {Ferrari}, F. 2019, \href{http://dx.doi.org/10.1093/mnras/stz2154}{\JournalTitle{\mnras}, 489, 1161}, \eprint arXiv:{1907.10188}

\bibitem[{Macci\'o {et~al.}(2008)Macci\'o \& Dutton \& Bosch}]{Maccio:2008pcd}
Macci\'o, A.~V., Dutton, A.~A., \& Bosch, F. C. v.~d. 2008, \href{http://dx.doi.org/10.1111/j.1365-2966.2008.14029.x}{\JournalTitle{\mnras}, 391, 1940}, \eprint arXiv:{0805.1926}

\bibitem[{{Mapelli} {et~al.}(2020){Mapelli} \& {Spera} {et~al.}}]{mapelli_2020}
{Mapelli}, M., {Spera}, M., {Montanari}, E., {et~al.} 2020, \href{http://dx.doi.org/10.3847/1538-4357/ab584d}{\JournalTitle{\apj}, 888, 76}, \eprint arXiv:{1909.01371}

\bibitem[{{Marinacci} {et~al.}(2025){Marinacci} \& {Baldi} {et~al.}}]{marinacci_2025}
{Marinacci}, F., {Baldi}, M., {Iorio}, G., {et~al.} 2025, \href{http://dx.doi.org/10.48550/arXiv.2510.06311}{\JournalTitle{arXiv e-prints}, arXiv:2510.06311}, \eprint arXiv:{2510.06311}

\bibitem[{Marra {et~al.}(2020)Marra \& Rodrigues \& de~Almeida}]{Marra:2020sts}
Marra, V., Rodrigues, D.~C., \& de~Almeida, {\'A}.~O. 2020, \JournalTitle{\mnras}, 494, 2875

\bibitem[{Meidt {et~al.}(2014)}]{Meidt:2014mqa}
Meidt, S.~E. {et~al.} 2014, \href{http://dx.doi.org/10.1088/0004-637X/788/2/144}{\JournalTitle{\apj}, 788, 144}, \eprint arXiv:{1402.5210}

\bibitem[{Mo {et~al.}(2010)Mo \& van~den Bosch \& White}]{MoBoschWhite2010}
Mo, H., van~den Bosch, F., \& White, S. 2010, {Galaxy Formation and Evolution} (Cambridge University Press)

\bibitem[{Navarro {et~al.}(1997)Navarro \& Frenk \& White}]{Navarro:1996gj}
Navarro, J.~F., Frenk, C.~S., \& White, S.~D. 1997, \href{http://dx.doi.org/10.1086/304888}{\JournalTitle{\apj}, 490, 493}, \eprint arXiv:{astro-ph/9611107}

\bibitem[{{Peng} {et~al.}(2002){Peng} \& {Ho} \& {Impey} \& {Rix}}]{galfit_peng2002}
{Peng}, C.~Y., {Ho}, L.~C., {Impey}, C.~D., \& {Rix}, H.-W. 2002, \href{http://dx.doi.org/10.1086/340952}{\JournalTitle{\aj}, 124, 266}, \eprint arXiv:{astro-ph/0204182}

\bibitem[{Perrin {et~al.}(2014)Perrin \& Sivaramakrishnan {et~al.}}]{perrin2014updated}
Perrin, M.~D., Sivaramakrishnan, A., Lajoie, C.-P., {et~al.} 2014, in Space telescopes and instrumentation 2014: optical, infrared, and millimeter wave, Vol. 9143, SPIE, 1174

\bibitem[{{Ruschel-Dutra} \& {Dall'Agnol De Oliveira}(2020)}]{2020zndo...3945237R}
{Ruschel-Dutra}, D. \& {Dall'Agnol De Oliveira}, B. 2020, {danielrd6/ifscube v1.0}

\bibitem[{{Ruschel-Dutra} {et~al.}(2021){Ruschel-Dutra} \& {Storchi-Bergmann} {et~al.}}]{2021MNRAS.507...74R}
{Ruschel-Dutra}, D., {Storchi-Bergmann}, T., {Schnorr-M{\"u}ller}, A., {et~al.} 2021, \href{http://dx.doi.org/10.1093/mnras/stab2058}{\JournalTitle{\mnras}, 507, 74}, \eprint arXiv:{2107.07635}

\bibitem[{{S{\'a}nchez-Alarc{\'o}n} {et~al.}(2023){S{\'a}nchez-Alarc{\'o}n} \& {Rom{\'a}n} {et~al.}}]{2023A&A...677A.117S}
{S{\'a}nchez-Alarc{\'o}n}, P.~M., {Rom{\'a}n}, J., {Knapen}, J.~H., {et~al.} 2023, \href{http://dx.doi.org/10.1051/0004-6361/202346719}{\JournalTitle{\aap}, 677, A117}, \eprint arXiv:{2307.02527}

\bibitem[{{Sgalletta} {et~al.}(2023){Sgalletta} \& {Iorio} {et~al.}}]{sgalletta_2023}
{Sgalletta}, C., {Iorio}, G., {Mapelli}, M., {et~al.} 2023, \href{http://dx.doi.org/10.1093/mnras/stad2768}{\JournalTitle{\mnras}, 526, 2210}, \eprint arXiv:{2305.04955}

\bibitem[{Singer \& Price(2016)}]{bayestar}
Singer, L.~P. \& Price, L.~R. 2016, \href{http://dx.doi.org/10.1103/PhysRevD.93.024013}{\JournalTitle{Phys. Rev. D}, 93, 024013}

\bibitem[{Singer {et~al.}(2016)Singer \& Chen {et~al.}}]{Singer_2016}
Singer, L.~P., Chen, H.-Y., Holz, D.~E., {et~al.} 2016, \href{http://dx.doi.org/10.3847/2041-8205/829/1/l15}{\JournalTitle{ApJ}, 829, L15}

\bibitem[{{Singer} {et~al.}(2016){Singer} \& {Chen} {et~al.}}]{Singer_supp}
{Singer}, L.~P., {Chen}, H.-Y., {Holz}, D.~E., {et~al.} 2016, \href{http://dx.doi.org/10.3847/0067-0049/226/1/10}{\JournalTitle{ApJS}, 226, 10}, \eprint arXiv:{1605.04242}

\bibitem[{{Spera} {et~al.}(2019){Spera} \& {Mapelli} \& {Giacobbo} \& {Trani} \& {Bressan} \& {Costa}}]{spera_2019}
{Spera}, M., {Mapelli}, M., {Giacobbo}, N., {et~al.} 2019, \href{http://dx.doi.org/10.1093/mnras/stz359}{\JournalTitle{\mnras}, 485, 889}, \eprint arXiv:{1809.04605}

\bibitem[{{Swaters} \& {Balcells}(2002)}]{2002A&A...390..863S}
{Swaters}, R.~A. \& {Balcells}, M. 2002, \href{http://dx.doi.org/10.1051/0004-6361:20020449}{\JournalTitle{\aap}, 390, 863}, \eprint arXiv:{astro-ph/0204526}

\bibitem[{{Swaters} {et~al.}(2009){Swaters} \& {Sancisi} \& {van Albada} \& {van der Hulst}}]{2009A&A...493..871S}
{Swaters}, R.~A., {Sancisi}, R., {van Albada}, T.~S., \& {van der Hulst}, J.~M. 2009, \href{http://dx.doi.org/10.1051/0004-6361:200810516}{\JournalTitle{\aap}, 493, 871}, \eprint arXiv:{0901.4222}

\bibitem[{{Taylor}(2005)}]{topcat}
{Taylor}, M.~B. 2005, in Astronomical Society of the Pacific Conference Series, Vol. 347, Astronomical Data Analysis Software and Systems XIV, ed. P.~{Shopbell}, M.~{Britton}, \& R.~{Ebert}, 29

\bibitem[{Tommy {et~al.}(2023)Tommy \& White \& Alexandra \& Maio \& gunvor \& Dyer}]{kdepy}
Tommy, White, S., Alexandra, {et~al.} 2023, tommyod/KDEpy: v1.1.8

\bibitem[{{Troja} {et~al.}(2022){Troja} \& {Fryer} {et~al.}}]{troja2022}
{Troja}, E., {Fryer}, C.~L., {O'Connor}, B., {et~al.} 2022, \href{http://dx.doi.org/10.1038/s41586-022-05327-3}{\JournalTitle{\nat}, 612, 228}, \eprint arXiv:{2209.03363}

\bibitem[{van~der Kruit \& Freeman(2011)}]{vanderKruit:2011vt}
van~der Kruit, P.~C. \& Freeman, K.~C. 2011, \href{http://dx.doi.org/10.1146/annurev-astro-083109-153241}{\JournalTitle{Ann. Rev. Astron. Astrophys.}, 49, 301}, \eprint arXiv:{1101.1771}

\bibitem[{Vazdekis {et~al.}(2016)Vazdekis \& Koleva \& Ricciardelli \& R{\"o}ck \& Falc{\'o}n-Barroso}]{vazdekis2016uv}
Vazdekis, A., Koleva, M., Ricciardelli, E., R{\"o}ck, B., \& Falc{\'o}n-Barroso, J. 2016, \JournalTitle{Monthly Notices of the Royal Astronomical Society}, 463, 3409

\bibitem[{{Vigna-G{\'o}mez} {et~al.}(2018){Vigna-G{\'o}mez} \& {Neijssel} {et~al.}}]{vigna_gomez_2018}
{Vigna-G{\'o}mez}, A., {Neijssel}, C.~J., {Stevenson}, S., {et~al.} 2018, \href{http://dx.doi.org/10.1093/mnras/sty2463}{\JournalTitle{\mnras}, 481, 4009}, \eprint arXiv:{1805.07974}

\bibitem[{{Yang} {et~al.}(2024){Yang} \& {Troja} {et~al.}}]{2024Natur.626..742Y}
{Yang}, Y.-H., {Troja}, E., {O'Connor}, B., {et~al.} 2024, \href{http://dx.doi.org/10.1038/s41586-023-06979-5}{\JournalTitle{\nat}, 626, 742}, \eprint arXiv:{2308.00638}

\end{thebibliography}
